\documentclass{aa}
\usepackage{graphicx,natbib}
\usepackage[varg]{txfonts}
\usepackage{afterpage}
\usepackage{color}
\usepackage{float}
\usepackage{upgreek}
\usepackage{soul}
\usepackage{hyperref}

\newcommand{\Porb}{\ensuremath{P_{\mathrm{orb}}}}
\newcommand{\Prot}{\ensuremath{P_{\mathrm{rot}}}}
\newcommand{\Bm}{\ensuremath{\langle B\rangle}}
\newcommand{\Bz}{\ensuremath{\langle B_z\rangle}}

\newcommand{\Bq}{\ensuremath{\langle B_\mathrm{q}\rangle}}
\newcommand{\vsi}{\ensuremath{v\,\sin i}}
\newcommand{\Teff}{\ensuremath{T_{\mathrm{eff}}}}
\newcommand{\kms}{km\,s$^{-1}$}
\newcommand{\Feline}{Fe~\textsc{ii}~$\uplambda\,6149.2$\,\AA}
\newcommand{\Zeeman}{\Delta\lambda_{\rm Z}}
\newcommand{\ew}{W_\lambda}
\newcommand{\RI}{R_I^{(2)}(\lambda_I)}
\newcommand{\vr}{\ensuremath{v_{\mathrm{r}}}}

\bibpunct[ ]{(}{)}{;}{a}{}{,}

\begin{document}

\title{Super-slowly rotating Ap (ssrAp) stars: \\ New spectroscopic
  observations} 

\author{G.~Mathys\inst{1}
  \and D.~L.~Holdsworth\inst{2,3}
  \and M.~Giarrusso\inst{4}  
  \and D.~W.~Kurtz\inst{5,6}
  \and G.~Catanzaro\inst{4}
  \and F.~Leone\inst{7,4}}

\institute{European Southern Observatory,
  Alonso de Cordova 3107, Vitacura, Santiago, Chile\\\email{gmathys@eso.org}
\and
South African Astronomical Observatory, PO Box 9, Observatory 7935,
Cape Town, South Africa
\and
School of Physics, Engineering and Technology, University of York,
Heslington, York YO10 5DD, UK
\and
INAF–Osservatorio Astrofisico di Catania, via S. Sofia 78, 95123 Catania, Italy
\and
Centre for Space Research, North-West University, Mahikeng 2745, South
Africa
\and
Jeremiah Horrocks Institute, University of Central Lancashire, Preston
PR1 2HE, UK
\and
Dipartimento di Fisica e Astronomia, Sezione Astrofisica, Universit\`a
di Catania, Via S. Sofia 78, I-95123 Catania, Italy}

\date{Received $\ldots$ / Accepted $\ldots$}

\titlerunning{Spectroscopy of ssrAp stars}

\abstract
{The rotation periods of Ap stars range over five to six orders of
  magnitude. The origin of
  their differentiation
  remains unknown.}
{We carry out a systematic study of the longest period Ap stars to
  gain insight into their properties.}
{We analyse newly obtained spectra of a sample of super-slowly
  rotating Ap (ssrAp) star candidates identified by a TESS photometric
  survey to confirm that their projected equatorial velocity \vsi\ is
  consistent with (very) long rotation periods, to obtain a first
  determination of their magnetic fields, and to test their binarity.}
{The value of \vsi\ in 16 of the 18 studied stars is low enough for them to
  have moderately to extremely long rotation periods. All stars but
  one are definitely magnetic; for five of them, the magnetic field
  was detected for the first time. Five new stars with resolved
  magnetically split lines were discovered. Five stars that were not
  previously known to be spectroscopic 
  binaries show radial velocity variations; in one of them, lines from
  both components are observed.}
{}


\maketitle

\section{Introduction}
\label{sec:intro}
The existence of a substantial population of stars with rotation
periods of months to centuries among early type stars with radiative
envelopes and large-scale organised magnetic fields has become
increasingly well established in the past decade
\citep{2017A&A...601A..14M,2018MNRAS.475.5144S}. They represent the 
extreme tail of a rotation rate distribution that spans five to six
orders 
of magnitude, from less than one day to centuries. The spectral
types of the magnetic stars with radiative 
envelopes range from early F to O. They include, in order of increasing
temperature, chemically peculiar F, A and B stars (Fp, Ap and Bp
stars, often referred to collectively as Ap stars), magnetic early B
stars (with spectral types ranging from B5 to B0), and magnetic O
stars. All three groups show a similar distribution of the rotation
periods \citep{2018MNRAS.475.5144S}. However, while there is evidence
for some loss of angular 
momentum due to magnetospheric braking on the main sequence for the
hotter two groups \citep{2019MNRAS.490..274S}, that the same
  mechanism can account for the occurrence of super-slow rotation in
  Ap stars has not been established yet. The identification of the
  relevant processes has until now 
been hampered by the lack of relevant 
observational constraints.
The need
for additional observations leading to better characterisation of
possible connections between rotation, magnetic field, and other
physical properties in order to guide theoretical developments is
clear, but the most relevant missing elements of information have not
been specifically identified yet. In these conditions, focusing on the
group of stars showing the most extreme slow rotation appears as one
of the most promising approaches to distinguish other properties
related to spin rate differentiation. 

Early knowledge of extremely slow rotation in Ap stars was strongly
biased towards strongly magnetic stars, since the identification of
long periods was mostly a by-product of the interest in studying stars
showing resolved magnetically split lines, whose magnetic fields can
be determined in a particularly complete, assumption-free and
model-independent manner (see below for more details). Accordingly,
systematic surveys were carried out to search for such stars
\citep{1971ApJ...164..309P,1997A&AS..123..353M,2008MNRAS.389..441F,2012MNRAS.420.2727E,2019ApJ...873L...5C},  
many of which were subsequently studied in greater detail \citep[see
  in particular][]{2017A&A...601A..14M,2022MNRAS.514.3485G}. By
  contrast, weakly magnetic Ap stars that potentially have (very) long
  periods received little attention. This difference introduced a
  potential bias in the knowledge of (very) slow rotation in Ap stars,
  at best preventing potential connections between magnetic properties
  and rotation rate from being investigated. 

To address this bias, we undertook a systematic search
\citep{2020A&A...639A..31M,2022A&A...660A..70M,2024A&A...683A.227M}
based on the photometric survey carried out with TESS (the Transiting
Exoplanet Survey Satellite; \citealt{2015JATIS...1a4003R})
that allowed us to identify 144 super-slowly rotating Ap star
candidates. These identifications are based on the assumption that Ap
stars that do not show photometric variations other than due to
pulsation over the duration of a TESS cycle (27\,d) have, with a high
probability, a rotation period $\Prot>50$\,d. The latter is, by
(arbitrary) definition, the lower limit of super slow rotation
\citep{2020pase.conf...35M}. However, spectroscopic observations at
sufficient resolution are necessary to confirm the photometric
identification of ssrAp stars. In particular, it is necessary to verify that the
Ap classification of the candidate is correct and that its spectral
lines do not show Doppler broadening inconsistent with rotation
periods of 50\,d or longer.

For about half of the TESS-identified
ssrAp star candidates, such confirmation is available in the
literature. For the other half, we undertook a systematic programme to
acquire a spectrum of each star at medium-high to high resolving power
($40\,000\lesssim R\lesssim115\,000$). In addition to peculiarity and
rotational (non)-broadening confirmation, these observations allow the
derivation of preliminary constraints on the magnetic fields of the
targets. The results of the analysis of spectra obtained as part of
this project for a first batch of 27 ssrAp star candidates were
presented by \citet{2024A&A...691A.186M}. Seventeen of these targets
proved to be Ap stars with 
projected equatorial velocities consistent with super slow rotation,
or at least with moderately long rotation periods ($20\,{\rm
  d}\lesssim\Prot\lesssim50$\,d). Misclassification was probably
responsible for most of the broader-lined stars.

The present paper is based on a second series of spectra that were
obtained more recently as part of this project. Again, a minority
among them shows line broadening inconsistent with super-slow
rotation. We postpone their discussion to a later work, and
focus here exclusively on the sharpest-lined Ap stars of the studied
sample. 

The observations analysed in this work are described in
Sect.~\ref{sec:obs}. The methods used to extract information from them
are introduced in Sect.~\ref{sec:analysis}. Section~\ref{sec:meas}
presents the results of our measurements, which are summarised in
Sect.~\ref{sec:summary}. Graphical illustration and tabular material
related to the data discussed in Sects.~\ref{sec:obs} and
\ref{sec:meas} are included in Appendix~\ref{sec:obs_meas}. The conclusion
(Sect.~\ref{sec:conc}) gives an overview of the current status of the
project and of the forthcoming steps that we plan to carry out. 

\section{Observations}
\label{sec:obs}
In this paper, we present the results of the analysis of 18 stars that
do not show broad spectral lines. New spectra have been obtained for 
17 of them. Fifteen of these 17 stars were identified as ssrAp star
candidates in our TESS-based survey
\citep{2020A&A...639A..31M,2022A&A...660A..70M,2024A&A...683A.227M}. The 
other two, HD~117290 and HD~143487,  
had been reported by \citet{2008MNRAS.389..441F} to
have resolved magnetically split lines, suggestive of (very) slow
rotation.

For 11 of the 15 ssrAp star candidates, prior to our study, there
  were no published measurements of the mean magnetic field modulus
  $\Bm$ or of the mean quadratic magnetic field $\Bq$, no value of the
  rotation period $\Prot$, and no determination of the projected equatorial
  velocity \vsi\ with a resolution sufficient to assess critically the
  occurrence of super-slow rotation. Measurements of the mean
  longitudinal magnetic field $\Bz$ of four of them had been obtained at a
  few epochs, providing insufficient time sampling to set meaningful
  constraints on the rotation period. While these measurements confirm
  that the stars are detectably magnetic, their sensitivity to the
  observational geometry limits their usability for
  characterisation of the intrinsic field strengths.

The four stars of the present study for which more information
  was already available in the literature are HD~8441, with
  determinations of $\Prot$, \vsi, and $\Bz$ (but with $\Bq$ below the
  detection threshold); HD~221568 (\Prot, \vsi, $\Bz$, and a quantity
  that is essentially $\Bq$, at one epoch), HD~76460 and HD~110274
  (both of them $\Bm$). More details, including the relevant
  references, are given in Sect.~\ref{sec:meas}.

We recorded new spectra of 17 stars. Lower resolution
  ($R\sim40\,000$ against $R\sim115\,000$) spectroscopic 
observations of four of them had already been discussed by
\citet{2024A&A...691A.186M}: HD~11187, HD~17330, HD~203922, and
BD+35~5094. In addition, we complement the study of HD~151860 from
\citet{2024A&A...691A.186M} with the analysis of a UVES (UV-Visual
Echelle Spectrograph) spectrum from the ESO (European Southern
Observatory) Archive.

The new observations presented here were obtained with either HARPS-N
(the High Accuracy Radial velocity Planet Searcher for the Northern
Hemisphere; \citealt{2012SPIE.8446E..1VC}) at the TNG (Telescopio
Nazionale {\em Galileo\/}) or SALT-HRS (the Southern African Large
Telescope High Resolution \'echelle Spectrograph;
\citealt{2010SPIE.7735E..4FB}). The former covers the wavelength range
3830--6930\,\AA\ with a resolving power $R\sim115\,000$; the latter was
used in High Resolution mode to achieve $R\sim45\,000$ from
3700\,\AA\ to 8900\,\AA. The data obtained with these instruments were
complemented with spectra recorded with UVES
(\citealt{2000SPIE.4008..534D}; $R\sim107\,000$ over 4960--7070\,\AA)
fed by UT2 (Unit Telescope 2) of the ESO VLT (Very Large Telescope),
HARPS (the High Accuracy Radial velocity Planet Searcher;
\citealt{2003Msngr.114...20M}; $R\sim115\,000$
over 3800--6900\,\AA) fed by the ESO 3.6-m telescope, and FEROS (the
Fiber-fed Extended Range Optical Spectrograph;
\citealt{1999Msngr..95....8K}; $R\sim48\,000$ over 3530--9220\,\AA)
fed by the ESO 2.2-m telescope. The reduced HARPS and UVES spectra
were retrieved from the ESO Archive. 

The two 2008 UVES spectra of
HD~110274 (JD~2\,454\,523 and 2\,454\,535), the 2008 UVES spectrum of
HD~117290 (JD~2\,454\,515), all four UVES spectra
of HD~143487, and the single UVES spectrum of HD~151860, were acquired
as long time series (up to $\sim$2\,h) of short integrations
($\sim$60--80\,s) to study line profile variations due to
pulsation. These individual exposures were averaged in the reduction
process, to produce the spectra analysed here. The FEROS spectra were
recorded by a team including DWK and GM
\citep{2008MNRAS.389..441F,2012MNRAS.420.2727E} as part of a
systematic survey of cool Ap stars from a list compiled by
\citet{1993PhDT.......269M}. Whenever the corresponding reduced
spectra were available in the ESO Archive, we used them for the
present study; otherwise, we used our original reductions. (All the
observation dates are fully specified in Table~\ref{tab:meas}. We have
uploaded all the reduced spectra to the CDS.)
  
For the 13 stars that were not
studied by \citet{2024A&A...691A.186M}, 30\,\AA\ long portions of the
spectra are shown in 
Figs.~\ref{fig:spec6150_1} to \ref{fig:spec6150_3}. The
format is similar to that of Figs.~A.1 to A.5 of
\citet{2024A&A...691A.186M}, with the stars sorted in order of
increasing effective temperature, so that the two sets of observations
can be easily compared. The main transitions responsible
for a number of lines are identified and the way in which their
intensities evolve along the temperature sequence can be traced, as
described by \citet{2024A&A...691A.186M}. Remarkable chemical
peculiarities can also be spotted, such as the large intensities of
the Nd~{\sc iii}~$\uplambda\,6145.1$\,\AA\ and Pr~{\sc
  iii}~$\uplambda\,6160.2$\,\AA\ lines in HD~143487 and
HD~138777, which are indicative of considerable overabundances of
these elements.  

The \Feline\ line is resolved into its magnetically split components
in 11 of the 18 studied stars, which allows their mean magnetic field
modulus $\Bm$ (that is, the line intensity weighted over the visible
stellar hemisphere of the modulus of the magnetic vector) to be
determined. This resolution is reported here for 
the first time for five stars: HD~119794, HD~138777, HD~192686,
HD~221568, and BD+52~3124. The presence of resolved magnetically split
lines in the other six stars had already been announced by
\citet{2008MNRAS.389..441F}, \citet{2012MNRAS.420.2727E}, and
\citet{2024A&A...691A.186M}. A blown-up portion of the spectra of
HD~151860 and HD~203932 including the \Feline\ line was shown in
Fig.~2 of \citet{2024A&A...691A.186M} (note that both stars were
mistakenly assigned the same TIC number in this figure; the correct
identification of HD~151860 is TIC~170419024).  Here we present the same
spectral region in
Fig.~\ref{fig:spec6149} for the remaining nine stars with resolved
magnetically split lines. 

\section{Analysis}
\label{sec:analysis}
We present the results of the analysis of 35 spectra of the 18 studied
stars. This includes the determination of the following parameters:
the radial velocity $\vr$ (except for some FEROS spectra whose
wavelength calibration is inadequate for this purpose), the mean
magnetic field modulus $\Bm$ (only for the 27 spectra in which the
\Feline\ line is resolved into its magnetically split components), 
the mean quadratic magnetic field $\Bq$ (the square root of the sum of
the mean square magnetic field modulus and of the mean square
longitudinal magnetic field; see \citealt{1995A&A...293..746M}), and
the upper limit of the projected equatorial velocity \vsi.

Knowledge of each of these parameters serves different
  purposes. The magnetic field is a fundamental property that
  contributes to the definition of the evolution of the star. Its
  consideration is of particular relevance for the the understanding
  of the processes that affect the stellar rotation, given its
  potential to affect the latter through magnetic
  braking. Furthermore, studying the field variability represents the
  most effective way to constrain reliably rotation periods of the
  order of years to decades. Indeed, the variation amplitudes are
  generally large compared to the measurement uncertainties, and
  $\Bm$ and $\Bq$ determinations over long time spans are less subject
  to systematic instrumental effects and instrument to instrument
  calibration differences than $\Bz$ or photometric measurements
  \citep[see][for more details]{2020pase.conf...35M}. Whenever $\Bm$
  can be determined, that is the most valuable 
  piece of information that we can extract from our analysis. For
  stars that do not show resolved magnetically split lines, $\Bq$ represents
  a suitable substitute that allows one to build a larger sample for
  statistical study of the distribution of the magnetic field
  strengths and of the possible correlations between magnetic field
  strength and rotation period. For spectra in which both $\Bm$ and
  $\Bq$ can be derived, comparison between the two values is valuable
  to characterise their overall connection so that statistical studies
  can be meaningfully carried out with a mix of both in the sample,
  depending on their respective availability.

The main motivation for determination of the upper limit of \vsi\ 
  is to identify in the sample those stars that can plausibly have
  periods $\Prot>50$\,d, hence that are worth re-observing at multiple
  epochs as part of the quest for characterisation and undestanding of
  super-slow rotation in Ap stars. A first estimate is adequate for
  this purpose, without requiring the additional effort for accurate
  untangling of the various contributions to the Doppler-like component
  of the line broadening (see below). The procedure by which $\Bq$
  is determined represents a convenient means to this effect. The
  consistency between the \vsi\ upper limit derived in this way and
  the difference between the observed width of the Fe~{\sc
    i}~$\lambda\,5434.5$\,\AA\ magnetic null line and its instrumental
  and thermal  
  broadening was illustrated in Fig.~4 of \cite{2024A&A...691A.186M}.

The possible existence of a connection between
  rotation and binarity among Ap stars may potentially reflect the
  existence of a merger channel of Ap star formation
  \citep{2009MNRAS.400L..71F,2010ARep...54..156T} that is at the same
  time responsible for super-slow rotation
  \citep{2017A&A...601A..14M}. Accordingly, multi-epoch determinations
  of the radial velocity of ssrAp stars is a sensible
  undertaking. Similar to the $\Bm$ and $\Bq$ measurements, the $\vr$
  values that we determine here often represent a first reference to
  which future values will be compared. More generally, their
  (very) sharp lines allow ssrAp stars to have their radial velocities
  determined with exquisite accuracy. This opens an opportunity to
  study some of the widest binaries containing an Ap component.

The determination of $\vr$ is straightforward. The methods applied for
determination of $\Bm$, $\Bq$, and $(\vsi)_{\rm max}$ were described
in detail by \citet{2024A&A...691A.186M}.

The value of the mean magnetic field modulus is computed from the
wavelength separation of the components of the \Feline\ doublet by
application of the formula
\begin{equation}
  \lambda_{\rm r}-\lambda_{\rm b}=g\,\Delta\lambda_{\rm Z}\,\Bm,
\label{eq:Bm}
\end{equation}
where $\lambda_{\rm r}$ and $\lambda_{\rm b}$ are the wavelengths of
the red and blue line components, respectively; $g=2.70$ is the Land\'e
factor of the split level of the transition; $\Delta\lambda_{\rm
  Z}=k\,\lambda_0^2$, with $k=4.67 \times 10^{-13}$\,\AA$^{-1}$\,G$^{-1}$;
and $\lambda_0=6149.258$\,\AA\ is the nominal wavelength of the
transition. The wavelengths are expressed in Angstr\"oms and the
magnetic field in Gauss. The wavelengths are measured by fitting
simultaneously a Gaussian to each of the 
two split components; a third Gaussian is added to the fit for the
blending line affecting the blue wing of the \Feline\ line when its
contribution can significantly impact the derived values of
$\lambda_b$ and $\lambda_r$. The uncertainties of the derived $\Bm$
values are estimated by comparing the analysed spectra with those of
stars well observed over a full rotation cycle (or at least over a
wide enough range of phases), for which the uncertainties are given by
the scatter of the individual measurements around a smooth variation
curve. This procedure was discussed more extensively by
\citet{1997A&AS..123..353M}. 

For determination of the mean quadratic magnetic field, the observed
line widths in natural light (Stokes $I$) are characterised by the
second-order 
moments $\RI$ of their profiles about their centre of gravity
$\lambda_I$. Following
\citet{2006A&A...453..699M}, a sample of lines are analysed to
untangle three contributions to their broadening by 
performing a multiple least-squares fit of the form
\begin{equation}
\RI=a_1\,{1\over5}\,{\lambda_0^2\over
  c^2}+a_2\,{3S_2+D_2\over4}\,\Zeeman^2+a_3\,\ew^2\,{\lambda_0^4\over c^4}\,.
\label{eq:Bq}
\end{equation}
In this equation, $\ew$ is the equivalent width of the line, and $S_2$ and
$D_2$ are atomic parameters characterising the Zeeman pattern of the
considered transition. For details about these parameters as well as
about the definition and actual measurement of
$\RI$, see \citet{2017A&A...601A..14M}. The fit coefficients $a_1$,
$a_2$, and $a_3$ are related the physical parameters of
interest in the present context. 

The third term on the right-hand side of Eq.~(\ref{eq:Bq}) represents
the intrinsic line width \citep{2006A&A...453..699M}. It does not
provide any useful information for our purpose, but it must be taken
into account to avoid overestimating the other two terms. Magnetic
broadening is accounted for by the second term; the mean quadratic
magnetic field is given by the $a_2$ fit coefficient:
$\Bq=\sqrt{a_2}$. Its formal uncertainty is computed from the standard 
deviation $\sigma(a_2)$, which corresponds to the line-to-line scatter
about the best fit regression. Finally, the first term on the
right-hand side of Eq.~(\ref{eq:Bq}) includes all the contributions to
the line width that have the same wavelength dependence as the Doppler
effect. This includes the rotational and thermal broadening, any other
Doppler broadening of stellar origin (for instance, due to
microturbulence or to stellar oscillations), and instrumental
broadening. Accordingly, an upper limit of the projected equatorial
velocity can be derived from consideration of the $a_1$ fit
coefficient:
\begin{equation}
  \vsi\leq(a_1-a_{\rm inst}-1.474\times10^{-3}\,\Teff)^{1/2}\,,
\label{eq:vsimax}
\end{equation}
where $\Teff$ is the effective temperature of the analysed star. In
this equation, the contribution of the instrumental profile is $a_{\rm
  inst}=40.02$\,km$^2$\,s$^{-2}$ (SALT-HRS), $a_{\rm
  inst}=35.17$\,km$^2$\,s$^{-2}$ (FEROS), $a_{\rm
  inst}=6.70$\,km$^2$\,s$^{-2}$ (UVES), or $a_{\rm
  inst}=6.13$\,km$^2$\,s$^{-2}$ (HARPS and HARPS-N; the 1.23\,\kms
value given by \citealt{2024A&A...691A.186M} was mistaken). As for
$\Bq$, we 
adopt $\sigma(a_1)$ as the formal uncertainty of the upper limit
  of $\vsi$. It reflects the
  line-to-line scatter in the regression analysis that is carried out
  to derive the values of $\Bq$ and $(\vsi)_{\rm max}$ but does not
  account for possible systematic errors that may affect the
  determination of these quantities.

The measurement procedures sketched above have been discussed more
extensively by \citet{2024A&A...691A.186M}. This reference also
includes several examples that illustrate the approach followed and
its potential shortcomings. The procedure that we use for
  determination of the mean quadratic magnetic field and of the upper
  limit of the projected equatorial velocity is best suited for study
  of a statistical sample of stars. Admittedly, detailed modelling of
  individual stars may yield more accurate results since, for
  instance, 
  the micoturbulence contribution to the Doppler line broadening can
  be untangled from other effects. But its application is much more
  time-consuming and accordingly inappropriate for the purpose of
  obtaining preliminary constraints on the properties of a rather
  large set of stars in order to select the most interesting targets
  for a full multi-epoch study. Once observations providing a good
  sampling of the rotation cycle of a given star have been acquired,
  it will be much more meaningful to compute a detailed model.  

\section{Measurement results}
\label{sec:meas}
The results of the new measurements reported in this paper are
summarised in Table~\ref{tab:meas}. Below, we discuss them on a
star-by-star basis.  

\subsection{TIC~32259138 (HD~138777)}
\label{sec:hd_138777}
The \Feline\ line is resolved into its magnetically split components
in the SALT-HRS spectrum of HD~138777 analysed here. From this
splitting, we derived a value $\Bm=4698$\,G of the mean magnetic field 
modulus. As in numerous other Ap stars with resolved magnetically
split lines, the blue component of \Feline\ is significantly affected
by an unidentified blending line. Accordingly, we estimate the
uncertainty of $\Bm$ to be of the order of 100\,G. Taking into account
the involved uncertainties, the ratio between
the value of the mean magnetic field modulus and that of the mean
longitudinal magnetic field reported by \citet{2017AstBu..72..391R},
$\Bz=2.1$\,kG, is within the typical range \citep[see Sect.~5.2
of][for details]{2017A&A...601A..14M}. 

The precision of the derived value of the mean quadratic field (from
the analysis of Fe~{\sc i} lines),
$\Bq=5600\pm300$\,G, ranks among the best ones achieved from
SALT-HRS spectra. The $\Bq$ determination did not reveal
any significant rotational broadening. This lends strong support to
the conclusion that HD~138777 is an ssrAp star. 

Within uncertainties, the value of the radial velocity that we
derived, $\vr=-43.72\pm0.12$\,\kms\ is consistent with the values
of \citet{2017AstBu..72..391R} ($\vr=-46.3\pm2.7$\,\kms) and of
\citet{2020AJ....160...83S} ($\vr=-47.7\pm2.9$\,\kms). Thus, there
is no indication of binarity for HD~138777. 

\subsection{TIC 88202438 (HD~192686)}
\label{sec:hd_192686}
\begin{figure}
  \centering
  \includegraphics[width=0.95\linewidth,angle=0]{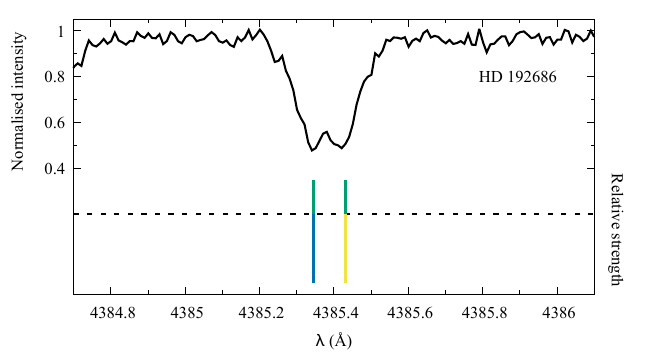}
\caption{Portion of the spectrum of HD~192686 showing the Fe~{\sc
    ii}~$\uplambda\,4385.4$\,\AA\ line. The Zeeman pattern of this line, a
  doublet, is illustrated below the spectrum. The amplitude of the
  splitting corresponds to the measured value of the mean magnetic
  field modulus, $\Bm=3573\,G$. The length of each vertical bar is
  proportional to the relative strength of the corresponding line
  component. The $\uppi$ components appear above the horizontal line
  (in green),
  the $\upsigma_+$ and $\upsigma_-$ components below it (in blue and
  yellow respectively).}
\label{fig:hd192686_4385}
\end{figure}

The hottest star of the present sample, HD~192686, was
observed with HARPS-N. The \Feline\ line is resolved into its
magnetically split components. Unusually, the red component is heavily
blended, by an unidentified line. This hampers the determination of
the mean magnetic field 
modulus from this line. Instead, we used the Fe~{\sc
  ii}~$\uplambda\,4385.387$\,\AA\  line to derive the value of this field
moment. This line is blend-free in HD~192686 (but in general not in
cooler stars). It arises from the transition b\,$^4{\rm
  P}_{1/2}$\,--\,z\,$^4{\rm D}^{\circ}_{1/2}$ while the transition
responsible for \Feline\ is b\,$^4{\rm D}_{1/2}$\,--\,z\,$^4{\rm
  P}^{\circ}_{1/2}$. Hence the two lines have essentially the same
Zeeman pattern, with a Land\'e factor $g$ of the $^4{\rm D}_{1/2}$
levels not significantly different from zero, and with $g=2.70$
(\Feline) and $g=2.68$ (Fe~{\sc ii}~$\uplambda\,4385.4$\,\AA) for the
$^4{\rm  P}_{1/2}$ levels. These lines are both pure doublets, in
which the two $\uppi$ components each are shifted from the line centre
by the same amount as each of the $\upsigma$ components. However, for
a given magnetic field strength, this shift is greater in the \Feline\
line than in the Fe~{\sc ii}~$\uplambda\,4385.4$\,\AA\ line, because of
the quadratic dependence of the Zeeman effect on the wavelength. The
observed splitting of Fe~{\sc ii}~$\uplambda\,4385.4$\,\AA\ in HD~192686
is illustrated in Fig.~\ref{fig:hd192686_4385}. From its
consideration, we derived a value 
$\Bm=3573 \pm 50$\,G for the mean magnetic field modulus.

Given the high effective temperature of HD~192686, Fe~{\sc ii} lines
were used to diagnose its mean quadratic magnetic field,
$\Bq=4730\pm230$\,G. Our analysis did not detect any significant
rotational broadening. This supports the
view that HD~192686 is an ssrAp star. The derived value of the radial
velocity, $\vr=-10.30\pm0.05$\,\kms, is consistent with the value
given in the Pulkovo compilation \citep{2006AstL...32..759G},
$\vr=-12.2\pm2.4$\,\kms.  

\subsection{TIC~93522454 (HD~143487)}
\label{sec:hd_143487}
The presence of resolved magnetically split lines in the spectrum of
HD~143487 was first reported by \citet{2008MNRAS.389..441F}. Here we
analyse the FEROS spectrum and the UVES spectrum that they obtained in
2007, two UVES spectra from 2008 and one from 2010, 
and a HARPS
spectrum from 2011, in addition to a recent (2023) SALT-HRS spectrum
that we recorded as part of the present project. 

The blend affecting the blue side of the \Feline\ line is
exceptionally strong: at some epochs, it is deeper than the Fe line
itself, in contrast to its appearance in the vast majority of Ap stars
in which the latter is resolved \citep[see Figs.~2 to 4
of][]{1997A&AS..123..353M}. This makes the determination of the mean
magnetic field modulus from the splitting of \Feline\ particularly
difficult. Unfortunately, due to the very high line density in the
spectrum of HD~143487, we could not identify any other Fe line with a
pure doublet or triplet Zeeman pattern that could be used to derive
the value of $\Bm$ in a similar approximation-free manner. As a
result, the estimated uncertainties of the 
reported values of this magnetic field moment are considerably higher
than in most other studied stars. However, the seven measurements that
we obtained, which range from $\Bm=4.00\pm0.10$\,kG to
$\Bm=4.47\pm0.10$\,kG, appear reasonably consistent with each other,
with possibly some indication of actual variability (see below). The
values from the first two epochs do not significantly differ from
those reported by \citet{2008MNRAS.389..441F}. The value of the mean
magnetic field modulus that we determine from the UVES spectrum of JD
2\,455\,374, $\Bm=4.20\pm0.06$\,kG, differs from that reported by
\citet{2013MNRAS.431.2808K}, $\Bm=4.75$\,kG. However, the latter was
derived from the consideration of lines of rare earth elements,
so that it is not directly comparable to our measurement based on an Fe
line. This is especially true given the presence of chemical
inhomogeneities on the surface of HD~143487, indicated by line intensity
variability (see below). The distribution of Fe and of the rare earth elements is likely
different, so that their lines do not sample the magnetic field in the
same way. 

Because of the high line density, the number of Fe~{\sc i} lines that
could be used to diagnose the mean quadratic magnetic field is much
smaller than usual, even though more severe blending was accepted in
the line selection. The challenge of finding suitable diagnostic lines
was compounded by the rather low S/N of most of the spectra. In
addition, for UVES, only lines with wavelengths
$\lambda\gtrsim4900$\,\AA\ are within the spectral range covered with
the configuration selected for the observation. However, the
availability of four spectra obtained at different epochs with this
instrument allowed us to use the same approach as had been adopted by
\citet{2024A&A...691A.186M} for TIC~167695608 to untangle more 
efficiently the contributions of the various broadening terms to the
observed line widths. Namely, we assumed that the intrinsic and
Doppler-like broadening terms do not vary with rotation phase to
average them over the four epochs, and only determined the magnetic
broadening term at each individual epoch, allowing better precision
to be achieved in the derived values of $\Bq$ \citep[see
Sect.~3.2.2 of][for details]{2024A&A...691A.186M}. 

Formally significant $\Bq$ values could be obtained from three of the
four UVES spectra and from the FEROS and SALT-HRS spectra. The value
obtained with HARPS is below the level of formal significance, and the
value derived from the fourth UVES spectrum marginally below this
level. This must, at least in part, be due to the comparatively low S/N
of the HARPS spectrum, to the limited number of suitable diagnostic
lines, and to the fact that many of them suffer an amount of blending
greater than in the other stars considered in this work. Given the
evidence for super-slow rotation of HD~143487,  
the upper limits of the projected equatorial velocity derived
from analysis of the various spectra, between $\vsi\lesssim5.8$\,\kms\  
and $\vsi\lesssim7.5$\,\kms, strongly suggest the occurrence of
significant crosstalk between the magnetic and Doppler terms of the
regression analysis performed to untangle them. This interpretation if
further supported by the fact that \citet{2013MNRAS.431.2808K} give a
value $\vsi=1.5$\,\kms\ for the projected equatorial velocity. In turn,
the occurrence of crosstalk probably accounts for the fact that the
low values of $\Bq$ determined 
in all cases are only at most marginally consistent
with the $\Bm$ values. Most likely, these $\Bq$ values are
underestimated because of crosstalk.

\begin{figure}
  \centering
  \includegraphics[width=1.0\linewidth,angle=0]{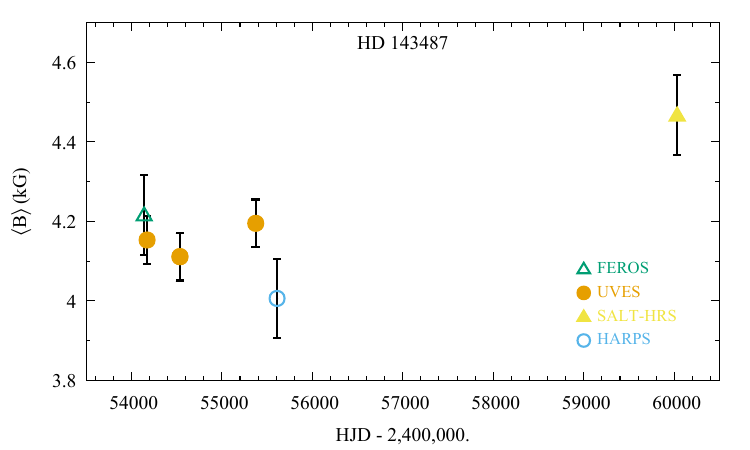}
\caption{Mean magnetic field modulus of HD~143487 against observation 
  date. Different
  point types and colours are used to distinguish the observations
  performed with different instruments.}
\label{fig:bm_hjd_143487}
\end{figure}

While HD~143487 was not identified as an ssrAp star candidate in our
TESS-based survey, probably because of instrumental noise,
\citet{2021MNRAS.506.1073H} report the absence of rotational signal in
their analysis of the TESS data for this star. The spectra obtained at
different epochs, illustrated in Fig.~\ref{fig:hd143487_6150}, suggest
that the intensities of lines of some ions vary over time scales of
years. The mean magnetic field modulus, despite the unusually large
uncertainty affecting its measured values, may also be showing an
actual long-term trend, with a minimum possibly occurring around
2011--2012 (JD$\sim$2,456,000.)
or somewhat later (see Fig.~\ref{fig:bm_hjd_143487}).

The wavelength calibration of the FEROS spectrum of HD~143487 is
inadequate to determine the radial velocity. The radial velocity
values derived from the observations obtained with the other
instruments at the six other epochs range from
$\vr=-25.2\pm1.0$\,\kms\ to $\vr=-14.0\pm0.8$\,\kms. The
variability that they reflect is consistent with the fact that
HD~143487 is a spectroscopic binary with an orbital period
$\Porb=1353\fd9$\,d whose centre of mass has a radial velocity
$\vr=-23.97$\,\kms \citep{2023A&A...674A..34G}.

HD~143487 is a known roAp star. \citet{2010MNRAS.404L.104E} discovered
low amplitude pulsations in radial velocity variations in an ensemble
of spectral lines with a period around
$10$\,min. \citet{2013MNRAS.431.2808K} obtained a time-resolved series
of 62 spectra over 1.4\,h with UVES and directly detected radial
velocity variations in lines of Nd\,\textsc{iii} and Pr\,\textsc{iii},
and in the core of H$\alpha$, with a period of 9.63\,min. There is only
one Sector of TESS data, S12, where the highest peak in the amplitude
spectrum has a frequency consistent with the frequency
of the radial velocity variations, although the significance of that
peak is marginal \citep{2021MNRAS.506.1073H}.

\subsection{TIC~163801263 (HD~203922)}
\label{sec:hd_203922}
A CAOS (Catania Astrophysical Observatory Spectropolarimeter;
\citealt{2016AJ....151..116L}) spectrum of HD~203922 was analysed in
\citet{2024A&A...691A.186M}. Both the mean magnetic field modulus
and the mean quadratic magnetic field could be determined. The
magnetically split components of the $\Bm$ diagnostic line, \Feline,
which were only marginally resolved in the CAOS spectrum, are now
fully resolved with HARPS-N, so that a much more precise value of the
mean field modulus can be derived here, $\langle B \rangle = 3503 \pm 30$\,G. The 
value of the quadratic field measurement, $\langle B_q \rangle = 3830 \pm 120$\,G  
has a formal uncertainty among the lowest achieved as part of this
project. Within the
uncertainties, neither the mean magnetic field modulus nor the mean
quadratic magnetic field show any significant change between the
epochs of the CAOS and HARPS-N observations, which are almost 2 years
apart. This lends plausibility to a very long rotation period, of the
order of several years. However, the upper limit derived here for the
projected equatorial velocity, $\vsi\lesssim3.7$\,\kms, while more
stringent than the value $\vsi\lesssim5.3$\,\kms\ of
\citet{2024A&A...691A.186M}, does not rule out a moderately long
rotation period. 

The radial velocity does not show any significant change either
between the two epochs of observation. The formal uncertainty achieved 
in its determination from the HARPS-N spectrum, 0.05\,\kms, is among the
lowest obtained within the framework of this project. 

\subsection{TIC~170419024 (HD~151860)}
\label{sec:hd_151860}
Here we complement the results from \citet{2024A&A...691A.186M}
on HD~151860 by the analysis of the UVES spectrum of 2010 that was
part of the study of \citet{2013MNRAS.431.2808K}. The value of the
mean magnetic field modulus derived by measuring the splitting of the
\Feline\ line in this spectrum, $\Bm=2265\pm40$\,G, is considerably
smaller than the value determined in the 2023 SALT-HRS spectrum of
\citet{2024A&A...691A.186M}, $\Bm=3356\pm100$\,G. This confirms
the conclusion that the relative amplitude of variation of the
magnetic field of HD~151860 is above average, inferred from
consideration of the change in the mean quadratic magnetic field
between a FEROS spectrum from 2008 and a SALT-HRS spectrum of
2023. The value from 2010 obtained here, $\Bq=2.58\pm0.17$\,kG does
not significantly differ from the 2008 value, $\Bq=2.51\pm0.28$\,kG,
but both are much lower than the 2023 value,
$\Bq=5.41\pm0.56$\,kG. The mean quadratic field may plausibly have
been varying monotonically from 2008 to 2023, which would indicate a
rotation period of the order of 25 years or more.

However, this conjecture needs to be confirmed by additional
observations, especially since the value of the projected equatorial
velocity derived by \citet{2013MNRAS.431.2808K}, $\vsi=4.5$\,\kms,
which is reasonably consistent with the upper limit that we determine
here, $\vsi\lesssim4.2\pm0.3$\,\kms, seems somewhat large for a period
of decades. (One can also note that the magnetic field strength of
2.5\,kG that \citeauthor{2013MNRAS.431.2808K} infer from spectrum
synthesis analysis is also compatible with our determinations of $\Bm$
and $\Bq$.)

The difference between the values of the radial velocity 
measured from the 2010 UVES spectrum ($\vr=3.45\pm0.07$\,\kms) and the
2023 SALT-HRS spectrum ($\vr=4.10\pm0.25$\,\kms) is marginal. Both
values are also consistent with the one from the Gaia DR2
\citep{2018A&A...616A...1G} ($\vr=3.67\pm0.39$\,\kms). Thus there is
no clear evidence of binarity for HD~151860, contrary to what was
mistakenly reported by \citet{2024A&A...691A.186M}
\citep[see][]{2025A&A...694C...7M}. 

\subsection{TIC~238659021 (HD~8441)}
\label{sec:hd_8441}
\citet{1958ApJS....3..141B} discovered the magnetic field of the well
studied Ap star HD~8441 and obtained a series of measurements of its
mean longitudinal component ranging from $-750$ to $+400$\,G. However,
the most precise determinations of this field moment to date,
achieved by \citet{2007A&A...475.1053A}, yielded a maximum absolute
value $|\Bz|_{\rm max}=157\pm18$\,G. This value is consistent with the
fact that the mean quadratic magnetic field is below the detection
limit in our analysis of a HARPS-N spectrum (using Fe~{\sc ii}
diagnostic lines).  Visual inspection of our HARPS-N spectrum does not
show any evidence of differential line broadening, contrary to the
case of BD+35~5094 (Sect.~\ref{sec:bd+35_5094}).  Based on this
result, by comparison with other 
stars of this study, we estimate that $\Bq$ must definitely be lower
than 1\,kG, which is inconsistent with mean longitudinal field values
significantly greater than 200--300\,G. The uncertainties of the
measurements of HD~8441 reported by \citet{1958ApJS....3..141B} must
have been underestimated. \citet{2012AstL...38..721T} also mention the
absence of any magnetic enhancement of the spectral lines, lending
further support to the conclusion that, while HD~8441 is definitely a
magnetic Ap star, it has one of the weakest fields in the class.

The upper limit of the projected equatorial velocity that we derive
from our analysis, $\vsi\lesssim3.1\pm0.2$\,\kms, is consistent with the
published value $\vsi\leq2.9\pm0.6$\,\kms
\citep{2002A&A...394..151C}. The rotation period of this ssrAp star,
$\Prot=69\fd51\pm0\fd01$, is well determined
\citep{2017PASP..129j4203P}. Using the radius value from the TIC
($R=5.16\,{\rm R}_{\odot}$), and $\Prot=69\fd51\pm0\fd01$, we find an
equatorial rotation velocity of $v_{\rm eq} = 2\uppi R/P =
3.76$\,km~s$^{-1}$. Coupled with our determination of
$\vsi\lesssim3.1\pm0.2$\,\kms, this yields $i\lesssim56^{\circ}$. This
limit is not very stringent. 

\citet{1998A&AS..130..223N} determined the orbital parameters of the
``short'' period ($\Porb=106\fd357$) within the HD~8441 triple
system. The epoch of our HARPS-N observation corresponds to orbital
phase $0.937\pm0.008$. The value that we derive for the radial velocity,
$\vr=3.68\pm0.03$\,\kms, appears consistent with the curve shown in
Fig.~1 of \citeauthor{1998A&AS..130..223N}

\subsection{TIC~251976407 (HD~221568)}
\label{sec:hd_221568}
\citet{1971ApJ...164..309P} determined a magnetic field strength of
1.8\,kG from the analysis of differential magnetic broadening of
spectral lines with narrow and broad Zeeman patterns in
HD~221568. The value that he derived has essentially the same
physical meaning as the mean quadratic magnetic field. In the spectrum
that he analysed, the lines were not resolved into their magnetically
split components.

By contrast, in our HARPS-N spectrum of HD~221568, the two components
of the \Feline\ line are resolved. The unidentified blending line on
the blue side of \Feline\ is separated enough from the latter so that
good precision is achievable in the determination of the
mean magnetic field modulus, $\Bm=2433\pm30$\,G. Although the
uncertainty of the field value obtained by
\citeauthor{1971ApJ...164..309P} is unknown, the difference between it
and the value of that we derive for the mean quadratic magnetic field,
$\Bq=2790\pm250$\,G, is probably significant and representative of the
actual variability of the observed field as a result of stellar
rotation. Unfortunately, we do not know the date on which the data
analysed by \citeauthor{1971ApJ...164..309P} were obtained, so that we
cannot estimate the phase difference between the two measurements,
even though the accuracy of the value of the rotation period,
$\Prot=159\fd10\pm0\fd03$ \citep{2017PASP..129j4203P}, is good enough
to derive a meaningful constraint. The occurrence of $\Bq$ variations
of rather large amplitude appears especially plausible considering the
variability of the spectral line intensities (summarised by
\citeauthor{2017PASP..129j4203P}) and the significant difference
between the values of the mean longitudinal magnetic field determined
at two epochs separated by 19\,d \citet{2018AstBu..73..178R}:
$\Bz=722\pm32$\,G and $\Bz=461\pm34$\,G.

Contrary to other stars for which the published value of the effective
temperature is similar, such as HD~89069 (Sect.~\ref{sec:hd_89069}),
the lines of 
Fe~{\sc i} in HD~221568 are weak or absent, so that the diagnostic
lines used for the present analysis pertain to Fe~{\sc ii}. 
Their rotational broadening is below the detection
limit, which is consistent with the long rotation period. The radial velocity
value at the epoch of our observation, $\vr=-8.33\pm0.12$\,\kms,
differs significantly from those published by
\citet{2018AstBu..73..178R}: $-6.9\pm2.7$\,\kms\ and
$-0.3\pm2.4$\,\kms. Thus, HD~221568 appears to be a spectroscopic
binary.  

\subsection{TIC~301918605 (HD~17330)}
\label{sec:hd17330}
The mean quadratic magnetic field of HD~17330 was below the detection
threshold in our previous analysis of a CAOS spectrum
\citep{2024A&A...691A.186M}. Here, based on a HARPS-N spectrum,
we achieve a $6.3\,\upsigma$ determination of the value of this field
moment, $\Bq=1130\pm180$\,G. We also derive a much more stringent
upper limit of the projected equatorial velocity,
$\vsi\lesssim1.4\pm0.1$\,\kms, which confirms that HD~17330 must be an
ssrAp star. The value of the radial velocity,
$\vr=-12.42\pm0.03$\,\kms, is the most precise one measured as part of
this study. While small, the difference between it and the
value obtained from the CAOS spectrum recorded 806\,d earlier,
$\vr=-13.10\pm0.12$\,\kms, is significant, consistent with reports
in the literature that HD~17330 is a spectroscopic binary
\citep{2024A&A...691A.186M}.

\subsection{TIC~341616734 (HD~89069)}
\label{sec:hd_89069}
\citet{2020AstBu..75..294R} discovered the magnetic field of
HD~89069. Their four measurements of the mean longitudinal magnetic
field \citep[see also][]{2017AstBu..72..391R} range from $\Bz=-720$\,G
to $\Bz=20$\,G, indicating variability on a time scale of days to
weeks. Analysis of two HARPS-N spectra of the star yield 
determinations of the mean quadratic magnetic field at the
$\sim$29\,$\upsigma$ and $\sim$36\,$\upsigma$ level. The precision of the
latter measurement, 90\,G, is the best we have ever achieved for any star. 
It appears to represent
the precision limit achievable with the observations that we are
performing. However, the derived value of $\Bq$ does not show any
significant change between the two epochs of observation, 23\,d
apart. 

Despite the relatively high derived value, $\Bq\simeq3.3$\,kG, the spectral
lines observed in the visible spectrum of HD~89069 are not resolved
into their magnetically split components at the epoch of our
observation, even though some of them 
(for instance, \Feline\ and Fe~{\sc i}~$\uplambda\,6336.8$\,\AA) show
clear magnetic 
broadening. This reflects the fact that magnetic resolution is smeared
out by a moderate amount of rotational broadening, consistent with
the upper limit of the projected equatorial velocity derived from our analysis,
$\vsi\lesssim4.3$\,\kms. Nevertheless, one should keep in mind that
depending on magnetic variability, which cannot be characterised yet,
some spectral lines (for instance, \Feline) may show at least partial
magnetic resolution over part of the stellar rotation cycle. 

\citet{1977IBVS.1288....1B} reported that HD~89069 is photometrically
variable. The time range spanned by their observations was too short
to determine the rotation period, but they showed an illustration
based on an assumed, arbitrary value $\Prot=18$\,d.
The TESS data show no evidence of such a period
\citep{2022A&A...660A..70M}. 

The star was mistakenly labelled as HD~86069 in Fig.~B.6 of
\citet{2022A&A...660A..70M}. There are now 120-s TESS
data available for Sectors~14, 20, 21, 26, 40, 41, 47, and 53, with
the earlier data reprocessed. An amplitude spectrum of the data
ensemble shows a hump of low-frequency amplitude, but no obvious
$\upalpha^2$\,CVn rotational peak. The amplitude spectrum out to a
frequency $5$\,d$^{-1}$ does not look significantly
different from Fig.~B.6 of
\citet{2022A&A...660A..70M}. Despite the reported variability of
the mean longitudinal magnetic field, no
rotational modulation is apparent in the TESS photometric
data. Nevertheless, the reported magnetic variability and the probable
rotational broadening of the spectral lines (see above) suggest that
HD~89069 is more likely a star with a moderately long period
($20\,{\rm d}\lesssim\Prot\lesssim50$\,d) than an ssrAp star. 

The measurements of \citet{2017AstBu..72..391R} and
\citet{2020AstBu..75..294R} 
suggest that the radial velocity of 
HD~89069 is variable, albeit close to the threshold of significance,
given the rather large uncertainties (up to 4.4\,\kms) of the derived
values. Our determinations, $\vr=-9.23\pm0.10$\,\kms\ and
$\vr=-9.17\pm0.05$\,\kms\ are within the range spanned by the published ones
($-12.7\,{\rm km\,s}^{-1}\lesssim\vr\lesssim3.8$\,\kms), but they do
not significantly differ from each other. The
binarity of HD~89069 needs to be confirmed by additional precise
radial velocity measurements.   

\begin{figure}
  \centering
  \includegraphics[width=1.0\linewidth,angle=0]{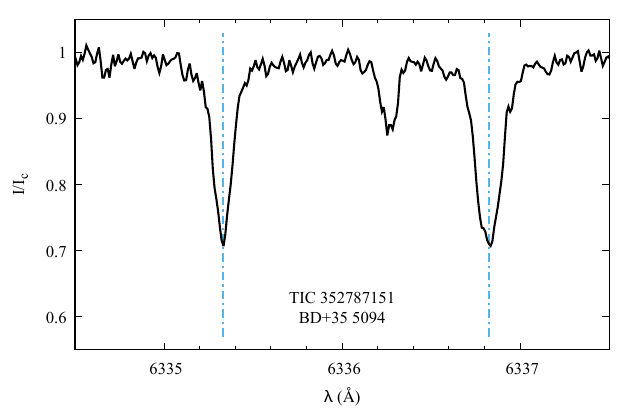}
  \caption{Comparison of the profiles of the Fe~{\sc
      i}~$\uplambda\,6335.3$\,\AA\ 
  and $\uplambda\,6336.8$\,\AA\ lines in BD+35~5094. The wavelengths are
  in the laboratory reference frame. The Zeeman patterns of the lines
  seen in the considered wavelength range were shown in Fig.~6 of
  \citet{2024A&A...691A.186M}. One can clearly see that Fe~{\sc
    i}~$\uplambda\,6336.8$\,\AA\ is significantly broader than Fe~{\sc
    i}~$\uplambda\,6335.3$\,\AA, which reveals the presence of a
    magnetic field in BD+35~5094.}
\label{fig:bdp35_5094_6336}
\end{figure}

\subsection{TIC~352787151 (BD+35~5094)}
\label{sec:bd+35_5094}
The mean quadratic field of BD+35~5094 was below the detection limit
in our analysis of a CAOS spectrum reported by
\citet{2024A&A...691A.186M}. We suspected the occurrence of some
crosstalk between the rotation and magnetic terms of the regression
carried out to untangle the various broadening contributions to the
observed line width. This suspicion was borne out by the rather high
value derived for the upper limit of the projected equatorial velocity
($\vsi\lesssim6.8\pm0.6$\,\kms). Its plausibility is strengthened by
the fact that we determine here a much lower value of this upper
limit, $\vsi\lesssim2.0\pm0.2$\,\kms, from the analysis of a HARPS-N
spectrum. In the latter, there does not appear to be any significant
crosstalk between the rotation and magnetic terms. Accordingly, it is
rather unexpected that the mean quadratic magnetic field remains below
the detection threshold.

This is all the more surprising since the differential broadening
in pairs of lines of different magnetic sensitivities is clearly seen
by visual inspection of the spectrum. This is illustrated in
Fig.~\ref{fig:bdp35_5094_6336}, where the additional broadening of the
Fe~{\sc i}~$\uplambda\,6336.8$\,\AA\ line with respect to the
neighbouring Fe~{\sc i}~$\uplambda\,6335.3$\,\AA\ line is clearly
visible. (This effect was already reported by
\citealt{2024A&A...691A.186M}.) The former line is one of the most
magnetically sensitive 
Fe~{\sc i} lines in the observed spectral range; the magnetic
sensitivity of the latter is much lower \citep[see][for more
details]{2024A&A...691A.186M}. Because the two lines occur at
neighbouring wavelengths and because their equivalent widths are similar, 
the difference between their widths can be
unambiguously attributed to magnetic broadening, albeit caused by a 
rather weak field. The weakness of this field accounts for the failure
to untangle its contribution from the intrinsic and rotational
terms in the regression analysis carried out to try to
determine its mean quadratic value. Based on consideration of the
results of the analysis of the stars of the sample for which the
lowest formally significant values of $\Bq$ are derived, we estimate
that the mean quadratic magnetic field of BD+35~5094 must be slightly
weaker 
than $\sim$1\,kG. Actually, the main contribution to the observed line
width is the intrinsic term. With the small rotational broadening (see
above), BD+35~5094 must be a weakly magnetic ssrAp star.

The value of the radial velocity derived from the HARPS-N spectrum,
$\vr=-6.81\pm0.05$\,\kms, differs significantly from that determined
from the CAOS observation obtained 2.2\,yr earlier,
$\vr=-9.44\pm0.12$\,\kms. Both values are also different from the one
published in the Gaia DR2 Catalogue \citep{2018A&A...616A...1G},
$\vr=-4.05\pm0.25$\,\kms. The ssrAp star BD+35~5094 belongs to a
spectroscopic binary system. 

\subsection{TIC~356088697 (HD 76460)}
\label{sec:hd_76460}
The resolution of the magnetically split components of the \Feline\
line in HD~76460 was first reported by \citet{2012MNRAS.420.2727E},
who derived a value $\Bm=3.6\pm0.2$\,kG, based on a FEROS spectrum
recorded in 2010 January. We analysed this spectrum, as well as a
SALT-HRS spectrum obtained in 2023 April. The value of the mean
magnetic field modulus that we determined from the latter,
$\Bm=3607\pm40$\,G, is not significantly
different from that of \citet{2012MNRAS.420.2727E}. The latter is also
fully consistent with the value that we derive by application of our
measurement procedure to a SALT-HRS spectrum: $\Bm=3586\pm50$\,G. 
The low estimated uncertainty of our determinations reflects the absence
of any major blending of the \Feline\ 
line, so that the measurement of its splitting is straightforward. The
fact that the mean magnetic field modulus does not appear to have
changed between two epochs of observation separated by more than 13
years suggests that the rotation period of HD~76460 may be extremely
long. This view is consistent with its identification as an ssrAp star
candidate \citep{2020A&A...639A..31M}, as well as with
the values determined for the upper limit of the projected equatorial
velocity.

Indeed, \citet{2012MNRAS.420.2727E} found $\vsi\lesssim3.0$\,\kms. 
As part of the analysis of the spectra for
determination of the mean quadratic magnetic field, we derived
$\vsi\lesssim4.0$\,\kms (FEROS) and $\vsi\lesssim3.9$\,\kms
(SALT-HRS). The values of the mean quadratic magnetic field determined 
at the two epochs, $\Bq=3890\pm260$\,G (2010) and $\Bq=3350\pm340$\,G
(2023), are consistent with each other within the uncertainties. If
any variation has occurred over the elapsed $\sim$13\,yr, it appears
moderate. This lends additional plausibility to the view that HD~76460
may have a rotation period of the order of decades. Both derived
values of the mean quadratic magnetic field are low with respect to
the mean field modulus, but not inconsistent with it. However, this
raises some concerns about the 
possible occurrence of crosstalk between the magnetic, Doppler and
intrinsic contributions in the multi-dimensional least-squares fit of
the observed second-order moments of the Stokes $I$ line profiles
about their centres \citep[see Sect.~3.2.1
of][]{2024A&A...691A.186M}. Nevertheless, critical evaluation of the 
analysis procedure does not show any clear indication of such
crosstalk. An intriguing possible alternative interpretation is that
HD~76460 has an unusual magnetic field structure. It would be
interesting to obtain spectropolarimetric observations to gain further
insight into this possibility. 

The wavelength calibration of the FEROS spectrum is inadequate for
determination of the radial velocity. Accordingly, we cannot assess
the possible binarity of HD~76460.

\subsection{TIC~369969602 (HD~128472)}
\label{sec:hd_128472}
Very little is known about HD~128472, which appears to be studied in
some detail for the first time here. We detected its magnetic field
from analysis of a SALT-HRS spectrum: $\Bq=1840\pm570$\,G. At the 
3.2\,$\upsigma$ level, this value is just above the threshold of formal
significance, but critical inspection of the dependences of the
various broadening contributions to the line profile width does not
suggest any significant crosstalk between them, so that the detection
can be regarded as firm. This is also supported by visual inspection
of the Fe~{\sc ii}~$\uplambda\,6335.3$\,\AA\ and $\uplambda\,6336.8$\,\AA\
lines, as the latter appears marginally broader than the former. On
the other hand, the upper limit of the projected equatorial 
velocity derived as part of the same analysis,
$\vsi\lesssim3.3$\,\kms, is consistent with super-slow rotation, but
a moderately long rotation period ($20\,{\rm
  d}\lesssim\Prot\lesssim50$\,d) cannot be definitely ruled out. Our
single measurement of the radial velocity of HD~128742 cannot be used
to assess its binarity since no other determination could be found
in the literature. 

\subsection{TIC~380607580 (HD~119794)}
\label{sec:hd_119794}
The SALT-HRS spectrum of HD~119794 shows resolved magnetically split
lines. The \Feline\ line suffers only minimal blending, so that the
estimated uncertainty on the derived value of the mean magnetic field
modulus is low: $\Bm=5369\pm40$\,G. We used Fe~{\sc ii} diagnostic
lines to determine the mean quadratic magnetic field,
$\Bq=4800\pm400$\,G. This value is low compared to that of
the mean field modulus. One may suspect that the value of $\Bq$ is
underestimated because of the occurrence of crosstalk between
the three terms contributing to the line broadening in the regression
analysis carried out to untangle them \citep[see][for
details]{2024A&A...691A.186M}. There is no unambiguous indication of
such crosstalk in the present case, but the upper limit derived for
the projected equatorial velocity, $\vsi\lesssim6.1\pm0.5$\,\kms,
seems somewhat high given the appearance of the magnetically split
components of the \Feline\ line. On the other hand, the Fe diagnostic
lines in HD~119794 are stronger than in most other
stars analysed in this work. Accordingly, the weak line
approximation underlying the method of determination of the mean
quadratic magnetic field \citep{2006A&A...453..699M} may be less
justified than in other cases. This may plausibly lead to
underestimating the value of $\Bq$.

Regardless, HD~119794 is definitely an Ap star with a sizeable
magnetic field that either rotates extremely slowly or must have a
moderately long rotation period. We could not find any published value
of its radial velocity in the literature, hence we cannot assess its
binarity from our single measurement. 

\subsection{TIC~403625657 (HD~11187)}
\label{sec:hd_11187}
The conclusion that ``HD~11187 is definitely a magnetic Ap star that
does not rotate particularly slowly'', based on the analysis of a CAOS
spectrum \citep{2024A&A...691A.186M}, was reached only shortly
before the star was re-observed with HARPS-N. Despite the higher
resolution and S/N of this latter spectrum, the mean quadratic
magnetic field remains below the detection threshold. The line
broadening is largely dominated by the rotational Doppler effect. The
upper limit of the projected equatorial velocity that we derived from
analysis of the HARPS-N spectrum, $\vsi\lesssim15.3\pm0.1$\,\kms, is
at most marginally lower than the CAOS-based
value, $\vsi\lesssim17.7\pm0.8$\,\kms. It remains compatible with a
rotation period of the order of days to weeks, as indicated by the
variability of the mean longitudinal magnetic field reported by the
group of the Special Astrophysical Observatory (Romanyuk, private
communication).

The values of the radial velocity determined from the CAOS
($\vr=5.61\pm0.27$\,\kms) and HARPS-N ($\vr=5.73\pm0.06$\,\kms)
spectra, obtained almost 2\,yr apart, do not significantly differ
from each other. There is no indication of binarity for HD~11187. 

\begin{figure}
  \centering
  \includegraphics[width=1.0\linewidth,angle=0]{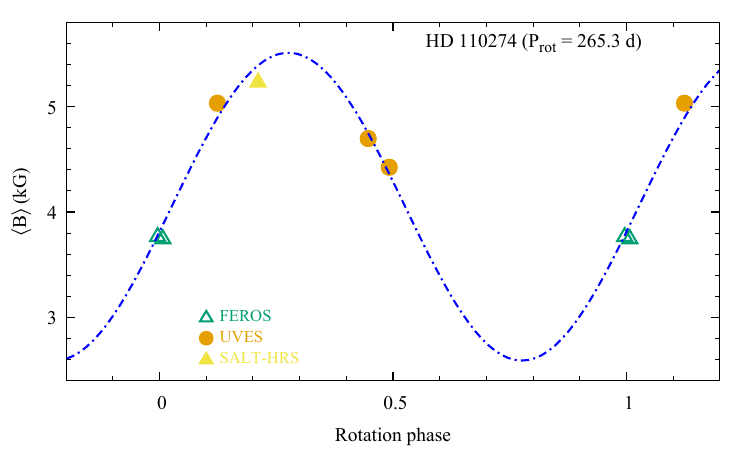}
  \includegraphics[width=1.0\linewidth,angle=0]{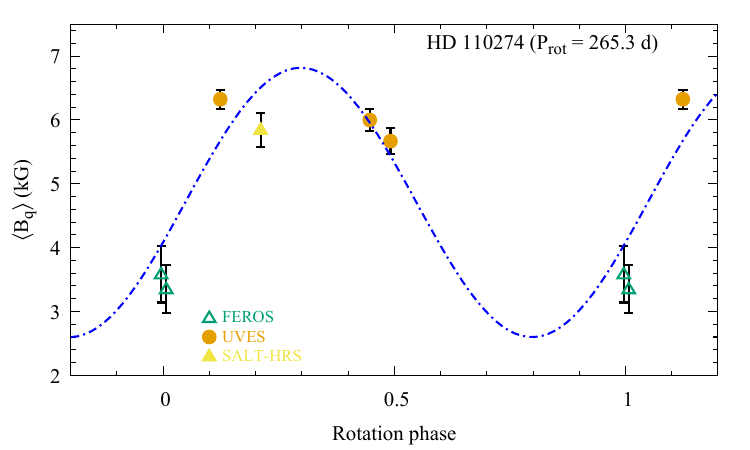}
\caption{Mean magnetic field modulus ({\em top\/}) and mean quadratic
  magnetic field ({\em bottom\/}) of HD~110274 against rotation phase
  computed with the value $\Prot=265\fd3$ of the rotation period
  derived by \citet{2008MNRAS.389..441F}. The phase origin has been
  set at JD~2\,454\,140.0. The dash-dotted blue curve represents the
  least-squares fit of the measurements by a sine curve. Different
  point types and colours are used to distinguish the observations
  performed with different instruments. The size of the error bars for
the $\Bm$ values does not exceed that of the symbols representing them.
}
\label{fig:bmq_phi_110274}
\end{figure}
 
\subsection{TIC~405516045 (HD~110274)}
\label{sec:hd_110274}
\citet{2008MNRAS.389..441F} discovered resolved magnetically split
lines in the spectrum of HD~110274. Their analysis was based on two
FEROS spectra and one UVES spectrum obtained at different epochs in
2007. The same group acquired UVES observations at two epochs in 2008,
which have not been used until now for magnetic field
diagnosis. We added a sixth epoch with a SALT-HRS spectrum recorded in
2023.

We performed determinations of the mean magnetic field modulus and of
the mean quadratic magnetic field from all six spectra. The values of
$\Bm$ that we derived from the FEROS spectra are fully consistent with
those reported by \citet{2008MNRAS.389..441F}. However, our analysis
of the 2007 UVES spectrum yielded $\Bm=5.25$\,kG, which is
significantly greater than the the value $\Bm=4.45$\,kG published by
\citeauthor{2008MNRAS.389..441F} from a measurement of the same
\Feline\ diagnostic line. We do not understand the origin of this
discrepancy, especially considering that the measurement is quite
straightforward as the \Feline\ line in HD~110274 suffers only minor
blending of its blue component by the unidentified line that severely
affects it in other Ap stars. Comparison of the line splitting with
that observed at other epochs strengthens our conviction that the
value of $\Bm$ that we derive from the UVES observation of 2007 is
correct and precise.

The values of $\Bm$ and $\Bq$ are plotted in
Fig.~\ref{fig:bmq_phi_110274} against the rotation phase computed
using the value of the period, $\Prot=265\fd3$, derived from
photometric observations by \citet{2008MNRAS.389..441F}. Both magnetic
field moments show variability with a relative amplitude above
average. They appear to vary nearly in phase with each other, which is
not unusual. (For comparison, see \citeauthor{2017A&A...601A..14M}
\citeyear{2017A&A...601A..14M}.) However, the phase coverage of the
measurements obtained until now is very incomplete, with a phase gap
of more than 0.5 cycle around the $\Bm$ and $\Bq$ minima. Accordingly,
the sine curve that is fitted to the data is only indicative; the
actual shape of the variation could significantly depart from
it. Observations at more epochs are needed to settle this issue.

For two of the UVES spectra, the Doppler contribution to the line
broadening is below the detection threshold. This is fully 
consistent with the length of the rotation period,
$\Prot=265\fd3$. 

Finally, we discovered significant variations of the radial velocity
of HD~110274. The wavelength calibration of the first epoch FEROS
spectrum is inadequate for radial velocity determination.  The values
determined at the five remaining epochs of
observation range from $\vr=-7.75\pm0.08$\,\kms\ to
$\vr=-1.43\pm0.10$\,\kms. This ssrAp star HD~110274 definitely
belongs to a spectroscopic binary, but the number of observations is
insufficient to constrain the orbital period. 

\subsection{TIC~419916333 (HD~117290)}
\label{sec:hd_117290}
The presence of resolved magnetically split lines in the spectrum of
HD~117290 was first reported by \citet{2008MNRAS.389..441F}. The
values of the mean magnetic field modulus that they derived from
measurement of two FEROS spectra and one UVES spectrum spanning one
month in 2007 do not show any significant variation. We reanalysed
these spectra together with another UVES observation (from 2008)
from the ESO Archive and a SALT-HRS spectrum that we recorded in
2023. The \Feline\ line is almost free from blends, so that the 
$\Bm$ values can be precisely determined. Those that we derived from
the 2007 spectra average at 6360\,G, consistent with the result
published by \citet{2008MNRAS.389..441F}. The more recent spectra
yield very different values: $\Bm=4509\pm30$\,G, in 2008, and
$\Bm=3020\pm40$\,G, in 2023. 

The ratio between the highest and lowest measured values of the mean
magnetic field modulus, $q\sim2.1$, is exceptionally large. Among the
Ap stars for which this value has been determined until now, it is exceeded
only by HD~9996 \citep[$q\sim3.3$,][]{2022MNRAS.514.3485G} and HD~57372
\citep[$q\sim3$,][]{2024A&A...687A.282H}, and of the same order as
HD~29578 \citep[$q\sim2.0$,][]{2022MNRAS.514.3485G} and HD~65339
\citep[$q\sim1.95$,][]{2017A&A...601A..14M}. This makes it probable
that the 2007 spectra were recorded in a phase interval around the
$\Bm$ maximum and the 2023 spectrum in the vicinity of the phase of
its minimum. If this speculation is correct, the $\sim$16\,yr
elapsed before the first and the most recent observations considered
here must approximately correspond to an odd number of half rotation
cycles. 

We did not identify HD~117290 as an ssrAp star candidate
as part of our TESS-based systematic survey. However, a lower limit of
the rotation 
period, $\Prot\gtrsim5.7$\,yr, had been derived from photometry
by \citet{2008MNRAS.389..441F}. Taking this into account, the star may
not have completed much more than 2.5 rotation cycles in 16\,yr. On
the other hand, the change undergone by the mean magnetic field
modulus from 2007 to 2008, $\sim$1.8\,kG, is close to half the
difference between the highest and lowest measured $\Bm$ value, which
does not appear compatible with a rotation period much longer than
5.7\,yr. Simultaneous consideration of these two constraints suggests
that the rotation period of HD~117290 must be of the order of
6\,yr. Admittedly, this conclusion relies on educated guesses that are
not proven, but it suggests that scheduling future observations so as
to test the viability of a $\sim$6\,yr period represents the strategy
with the highest chances of success.

The line density is very high in the spectrum of HD~117290, which
limits the number of Fe~{\sc i} lines that can be selected to diagnose
the mean quadratic magnetic field, even relaxing the blend-free
requirement applied to most other stars of this study. Furthermore,
the star undergoes considerable spectral line intensity variations
along its rotation cycle, so that different sets of diagnostic lines
must be used at different phases. The variability is illustrated in
Fig.~\ref{fig:hd117290_6150}, which shows the same spectral range as
Figs.~\ref{fig:spec6150_1} to \ref{fig:spec6150_3} as observed at
different epochs and with different instruments in HD~117290. In this
range, variability is observed mainly in the lines of Cr, Nd and
Pr, which all appear stronger when the magnetic field is weaker. In
other wavelength regions, identification of the variable lines 
is not straightforward and requires detailed analysis that is beyond
the scope of the present work. 

Similar to the mean magnetic field modulus, and apparently in phase
with it, the mean quadratic magnetic field shows variations of large
amplitude. The derived values of the two field moments are mostly
consistent with each other, albeit only marginally so in 2023, when
$\Bq<\Bm$. This 
formal discrepancy is well within the error margin, but may in part at
least result from the occurrence of a small amount of crosstalk
between the rotational and magnetic contributions to the line
broadening. The latter might also explain why a formally significant
upper limit of the projected equatorial velocity,
$\vsi\lesssim2.3\pm0.6$\,\kms\ is derived (which is compatible with a
rotation period of years) while rotational broadening is below the
detection limit in the other analysed spectra.

Finally, radial velocity variations, indicative of binarity, are
definitely detected. The wavelength calibration of the first-epoch
FEROS spectrum is inadequate for radial velocity determination, but
the values derived for the later four epochs range from
$\vr=-36.97\pm0.20$\,\kms\ to $\vr=-4.47\pm0.15$\,\kms. The fact that
the two values measured one month apart in 2007 hardly differ from
each other makes a long orbital period (of the order of years)
plausible, but this is not conclusive. Further observations are
required to obtain more meaningful constraints.

\begin{figure}
  \centering
  \includegraphics[width=1.0\linewidth,angle=0]{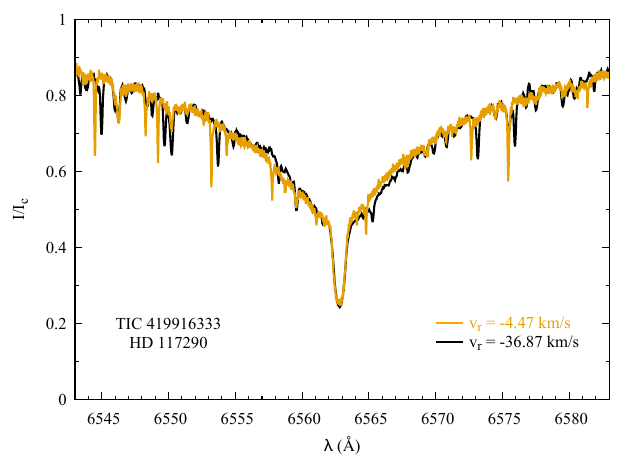}
  \caption{Spectra of HD~117290 at the epochs of the lowest
    and highest radial velocity values that were recorded in this
    study, $\vr=-4.47$\,\kms\ (2008 UVES spectrum) and
    $\vr=-36.87$\,\kms\ (2023 SALT-HRS spectrum). If lines of the two
    components of the binary are visible, this is when the
    greatest wavelength separation between them
    has been observed. The laboratory
    reference frame was used for the wavelengths of the lines of the
    sharp-lined Ap component. In this reference frame, the blue wing of
    the H$\alpha$ line is less depressed in the SALT-HRS spectrum than
    in the UVES one, and conversely, its red wing is more depressed in
    the SALT-HRS spectrum than in the UVES one. This is consistent
    with the presence of a (very) broad H$\alpha$ line from the
    secondary that is blue shifted in 2008 and redshifted in 2023 with
    respect to the H$\alpha$ line of the Ap star, which has a much
    narrower core. (The
    pairs of very narrow lines that appear shifted with respect to each
    other between the two spectra are of telluric origin.)
    }
 \label{fig:hd_117290_halpha}
\end{figure}

TESS photometry shows a rich $\updelta$~Sct frequency spectrum, which
is unusual for an Ap star. One may suspect the pulsation signal to
arise from the secondary of the binary system. As the $\updelta$~Sct
frequencies extend up to 40\,d$^{-1}$, this component is unlikely to
be cooler than $\Teff\sim7300$\,K, the value appearing in the TIC for
HD~117290. Thus, its contribution should be visible in the observed
spectrum. It is not apparent in the metal lines, probably because of
rapid rotation. This interpretation is borne out by consideration of
Fig.~\ref{fig:hd_117290_halpha}, in which one can distinguish a very
broad H$\alpha$ profile that undergoes a wavelength shift evolving
with time with respect to the sharp-cored H$\alpha$ line of the Ap
star. Thus, HD~117290 is a newly discovered double-lined spectroscopic
binary (SB2) system consisting of an ssrAp star and a fast-rotating
$\updelta$~Sct star. 

\subsection{TIC 430355895 (BD+52 3124)}
\label{sec:bd+52_3124}
There is very little information in the literature about BD+52~3124
besides spectral classification. In the HARPS-N spectrum of this star,
the \Feline\ line is resolved into its components by the magnetic
field. Their splitting indicates a value $\Bm=4237\pm50$\,G for the mean 
magnetic field modulus. The estimated uncertainty, 50\,G, reflects the
presence of a moderate blend affecting the blue component.

This is one of the few stars of the sample that is hot enough so that
the lines of Fe~{\sc ii} are better suited than those of Fe~{\sc i}
for the diagnosis of the mean quadratic magnetic field. The value of
the latter, $\Bq=4920\pm170$\,G, is fully consistent with that of the mean
magnetic field modulus. The value derived for the upper limit of the
projected equatorial velocity, $\vsi\lesssim3.2$\,\kms, is compatible 
with super-slow rotation, but it does not definitely exclude a
moderately long rotation period ($20\,{\rm
  d}\lesssim\Prot\lesssim50$\,d). We did not find any published
value of the radial velocity, so that we cannot assess the binarity of
BD+52~3124 on the basis of our single \vr\ measurement. 

\subsection{TIC~470837956 (BD+61~2565)}
\label{sec:bd+61_2565}
The analysis of the HARPS-N spectrum of BD+61~2565 reveals the
presence of a rather strong magnetic field: $\Bq=1680\pm140$\,G, which
is detected at the $12\,\upsigma$ level. This implies that the mean 
magnetic field modulus is below the threshold of resolution of the
magnetically split components of the \Feline\ line, $\sim$1.7\,kG
\citep{1997A&AS..123..353M}. However, the differential broadening of
lines of different magnetic sensitivities is clearly seen upon visual
inspection. The
derived value of the upper limit of 
the projected equatorial velocity, $\vsi\lesssim3.3$\,\kms, is
consistent with super-slow rotation, or possibly a moderately long
rotation period ($20\,{\rm d}\lesssim\Prot\lesssim50$\,d).

\begin{table}
  \scriptsize
  \caption{Properties of the studied stars.}
  \label{tab:properties}
  \centering
  \begin{tabular}{cccccc}
    \hline\hline\\[-4pt]
    TIC&Other ID&Resolved&Magnetic&Binarity&$\Prot$\\
       &        &lines   &field   &        &\\[4pt]
    \hline\\[-4pt]
    \multicolumn{6}{c}{ssrAp stars}\\[4pt]
    \hline\\[-4pt]
 32259138&HD~138777      &x&x&   &\\
 88202438&HD~192686      &x&x&   &\\
 93522454&HD~143487      &x&x&SB1&\\ 
170419024&HD~151860      &x&x&   \\
238659021&HD~8441        & &x&SB1&$69\fd51$\\
251976407&HD~221568      &x&x&SB1&$159\fd10$\\ 
301918605&HD~17330       & &x&SB1\\
356088697&HD~76460       &x&x&   \\
405516045&HD~110274      &x&x&SB1&$265\fd3$\\
419916333&HD~117290      &x&x&SB2&$\sim$6\,yr?\\[4pt]    
    \hline\\[-4pt]
    \multicolumn{6}{c}{Long period Ap stars}\\[4pt]
    \hline\\[-4pt]
163801263&HD~203922      &x&x&   &$\gtrsim25$\,yr?\\
352787151&BD+35~5094     & & &SB1\\
369969602&HD~128472      & &x&   \\
380607580&HD~119794      &x&x&   \\
430355895&BD+52~3124     &x&x&   \\
470837956&BD+61~2565     & &x&SB1\\[4pt]
    \hline\\[-4pt]
    \multicolumn{6}{c}{Moderately long period Ap star}\\[4pt]
    \hline\\[-4pt]
341616734&HD~89069       & &x&SB1\rlap{?}\\[4pt]
    \hline\\[-4pt]
    \multicolumn{6}{c}{Short period Ap star}\\[4pt]
    \hline\\[-4pt]
403625657&HD~11187      & &x&   \\[4pt]
    \hline
  \end{tabular}
    \tablefoot{The table is divided in
    four segments. The top one includes those stars for which the
    analysis presented in Sect.~\ref{sec:meas} indicates that
    they are in all probability ssrAp stars. The spectroscopic
    properties of the stars listed in the second segment are
    consistent with their having either moderately or extremely long
    rotation periods. The rotation period of the star from the third
  segment probably is moderately long ($20\,{\rm
    d}\lesssim\Prot\lesssim50$\,d). The fourth segment contains the
  only magnetic Ap star of this study to have a rotation period of a
  few days. In each segment, the
  stars are listed in order of increasing TIC number (Col.~1); an
  alternative ID is given is Col.~2. The visible range spectrum of
  those stars for which `x' appears in Col.~3 shows resolved
  magnetically split lines; the `x' flag in Col.~4 identifies those
  stars for which a significant detection of the magnetic field was
  achieved and at least one field moment could be determined, either
  in this study or in the literature (see text). Column~5
  indicates which 
  stars are spectroscopic binaries, distinguishing to the extent possible
  the single-lined (SB1) and double-lined systems. Finally, in Col.~6,
  we list the value of the rotation period, when available from the
  literature (the references are given in the relevant sub-sections of
Sect.~\ref{sec:meas}) or an estimated value or lower limit derived
as part of the present study.}
  \end{table}

The radial velocity at the epoch of the HARPS-N observation,
$\vr=-17.89\pm0.06$\,\kms, differs considerably from the published
value, $-49$\,\kms\ \citep{1963JO.....46..243B}. While the unspecified
precision of this determination, based on an objective prism spectrum,
must be limited, consideration of, for instance, Fig.~35 of this reference
supports the view that the change between the two epochs is highly
significant, hence
indicative of previously unreported binarity. Some of the sharpest
lines visible in the HARPS-N spectrum, including a couple of the
Fe~{\sc i} lines that we used to diagnose $\Bq$, have depths exceeding
0.9: the companion star does 
not contribute significantly to the observed spectrum. It must be much
fainter than the Ap component.

\section{Summary}
\label{sec:summary}
Table~\ref{tab:properties}, which is similar to Table~1 of
\cite{2024A&A...691A.186M}, summarises the results of our
analysis. For ten of the studied stars, the projected equatorial
velocity appears definitely consistent with super slow rotation. Six
are ssrAp star candidates for which the conclusion drawn from our TESS
survey receives here independent confirmation. A seventh one
(HD~17330) had already 
been confirmed by \citet{2024A&A...691A.186M}; for the remaining
three, values of the rotation period $\Prot>50$\,d\ were 
available in the literature. For six more stars (including two already
considered by \citealt{2024A&A...691A.186M}), the derived upper
limit of \vsi\ could indicate either super slow rotation or a
moderately long period ($20\,{\rm d}\lesssim\Prot\lesssim50$\,d),
while HD~89069 almost certainly has a moderately long rotation
period. While one could in principle compute minimum values of
  the rotation period of each star by combining the derived upper
  limit of \vsi\ with a radius value (available from the literature),
  we prefer to avoid this to take also into account our assessment of
  the reliability of $(\vsi)_{\rm max}$ -- for instance, to which
  extent it may be affected by crosstalk between the various line
  broadening terms in the regression analysis. Even though this
  approach involves a degree of subjectivity, it reduces the risk of
  overinterpretation of unqualified numerical values. As previously
  mentioned, we prefer to postpone the accurate determination of \vsi\
  and of a corresponding limit of \Prot\ to a future detailed analysis
  of stars for which multi-epoch observations confirm without
  ambiguity the occurrence of super-slow rotation.

Seventeen of the 18 stars of this study are definitely magnetic. For
five of them, the
detection is achieved here for the first time. No formally significant
value of the magnetic field could be determined for the eighteenth
star, BD+35~5094, but its spectrum is indicative of differential
broadening of lines of different magnetic sensitivity. For two of
  the stars, HD~8441 and HD~11187, the presence of the magnetic field
is inferred from mean 
longitudinal magnetic field measurements from the
literature. Formally significant values of the mean quadratic
  magnetic field were derived for the remaining 15. The
occurrence of
resolved magnetically split lines in the visible spectrum of five
stars is reported here for the first time. In total, eleven of the 18 
stars of this study show magnetic resolution of the \Feline\ line,
making it possible to determine their mean magnetic field
  modulus. 

At least eight, and probably nine of the analysed stars are
spectroscopic binaries. The binarity of five of them does not seem to
have been reported before (although, for BD+35~5094, different radial
velocity values had been published). The binarity of HD~89069 was
suspected, but it needs to be confirmed through additional precise
measurements.

\section{Conclusion}
\label{sec:conc}
The main purpose of the study presented by
\citet{2024A&A...691A.186M} and in this paper was to obtain
spectroscopic confirmation of the viability of the ssrAp star
candidates identified photometrically by
\citet{2020A&A...639A..31M,2022A&A...660A..70M,2024A&A...683A.227M},
for which no relevant information about the rotation period or the
projected equatorial velocity was found in the literature. We have now
completed the acquisition of the spectra of all but two such stars
with magnitude $V\leq9.5$. Upon visual inspection, a fraction of them
definitely show rotational line broadening incompatible with values
of the rotation period $\Prot>50$\,d. We provisionally left them
aside, postponing to a later stage the investigation of the reasons
why no photometric variability on rotational time scales shorter than
$\sim$20\,d were found in our TESS survey. The analysis of the
spectra obtained as part of our project for the remaining ssrAp star
candidates is now complete. Not only have we determined upper limits
of the projected equatorial velocities of the observed stars, but also
we have performed first measurements of their magnetic fields and, in a
number of cases, we have been able to test their possible binarity. The
outcome represents a valuable basis to derive constraints on the
distribution of the rotation rates and on the connections between slow
rotation and other physical parameters.

Nevertheless, it would be premature at this stage to undertake
such an endeavour, because there is still both in our files and in
observatory archives a wealth of unexploited spectroscopic material
available about those stars for which information supporting the
identification as ssrAp star candidates exists in the
literature. For many of these stars, the existing observational
material can be used to extract constraints about the rotational
velocity, the magnetic field, and/or the binarity, that have not been
considered until now. By carrying out this task as the next step of
our project, we shall build a more exhaustive database that will 
put us in the position to explore the connections between long
rotation periods and other stellar properties in a more statistically
significant manner than if we limited ourselves to the data obtained
by us that we have analysed until now. 

On the other hand,
the 16 stars listed in Table~\ref{tab:properties} under `ssrAp stars'
or `Long period Ap stars' fulfil the necessary condition of low enough
projected equatorial velocity to be ssrAp stars. However, only three
of them have been confirmed until now to have rotation periods
$\Prot\gtrsim50$\,d. Determination of the periods of the remaining 13
is required to ascertain their rotational status. In most cases, this
will be best done by studying the variability of their magnetic field
\citep{2020pase.conf...35M}. The most suitable strategies to this
effect have been discussed in detail by
\citet{2024A&A...691A.186M}. Their implementation, which represents
one of the next steps of our project, is already in progress.

\begin{acknowledgements}
Based on observations made with the Italian
Telescopio Nazionale 
{\it Galileo\/} (TNG) operated on the island of La Palma by the Fundación
{\it Galileo Galilei\/} of the INAF (Istituto Nazionale di Astrofisica) at the
Spanish Observatorio del Roque de los Muchachos of the Instituto de
Astrofisica de Canarias; on observations obtained with the Southern
African Large Telescope under the proposal codes
2022-2-SCI-017, 2024-1-SCI-007, 2024-2-SCI-023 (PI: Holdsworth); on
observations collected at the European 
Southern Observatory under ESO programme 078.D-0080; and on processed
data obtained from the ESO Science Archive Facility. This work
has made use of the VALD database, operated at Uppsala 
University, the Institute of Astronomy RAS in Moscow, and the
University of Vienna. This research has also made use of the SIMBAD
database, operated at CDS, Strasbourg, France. 
\end{acknowledgements}

\bibliographystyle{aa}
\bibliography{ssrAp2}

\onecolumn 
  
\begin{appendix}

\section{Observations and measurements}
\label{sec:obs_meas}

\begin{table*}[h!]
  \scriptsize
  \caption{Measurements of the spectra of the ssrAp star
    candidates.}
\label{tab:meas}
\setlength{\tabcolsep}{4pt}
\begin{tabular*}{\textwidth}[]{@{}@{\extracolsep{\fill}}rcrccrrrrrrcrrcrr}
\hline\hline\\[-4pt]
    \multicolumn{1}{c}{TIC}&Other ID&\multicolumn{1}{c}{$V$}&HJD&Instr&\multicolumn{1}{c}{S/N}&\multicolumn{1}{c}{\Teff}&\multicolumn{1}{c}{$v_{\rm r}$}&\multicolumn{1}{c}{$\sigma(v_{\rm r})$}&\multicolumn{1}{c}{$\Bm$}&\multicolumn{1}{c}{$\sigma(\Bm)$}&$N_{\rm g}$&\multicolumn{1}{c}{$\Bq$}&\multicolumn{1}{c}{$\sigma(\Bq)$}&$N_l$&\multicolumn{1}{c}{$(v\,\sin\,i)_{\rm max}$}&\multicolumn{1}{c}{$\sigma(v\,\sin\,i)$}\\
    &&&$(2,400,000.+)$&&&\multicolumn{1}{c}{(K)}&\multicolumn{1}{c}{(\kms)}&\multicolumn{1}{c}{(\kms)}&\multicolumn{1}{c}{(G)}&\multicolumn{1}{c}{(G)}&&\multicolumn{1}{c}{(G)}&\multicolumn{1}{c}{(G)}&&\multicolumn{1}{c}{(\kms)}&\multicolumn{1}{c}{(\kms)}\\[4pt]
\hline\\[-4pt]
 32259138&HD 138777      & 9.73&60432.570&S&150& 7352&-43.72&0.12& 4698&100&3& 5600& 300& 46&  --&\\[3pt]
 88202438&HD 192686      & 8.88&60540.457&N& 85&11670&-10.30&0.05& 3573& 50&2& 4730& 230& 64&  --&\\[3pt]
 93522454&HD 143487      & 9.43&54140.822&F&170& 7114&      &    & 4216&100&3& 3940& 780& 21& 5.8&0.8\\
         &               &     &54173.913&U&180&     &-23.36&0.69& 4153& 60&3& 2630& 920& 17& 7.5&1.4\\
         &               &     &54535.850&U&135&     &-14.00&0.79& 4111& 60&3& 2890& 750& 17& 7.5&1.4\\
         &               &     &54536.815&U&225&     &-14.24&0.71& 4111& 60&3& 3280& 630& 17& 7.5&1.4\\
         &               &     &55374.730&U&325&     &-25.20&0.99& 4195& 60&3& 3330& 760& 14& 7.5&1.4\\
         &               &     &55609.836&H& 50&     &-22.27&0.47& 4006&100&3&   --&    & 19& 6.9&1.3\\
         &               &     &60032.483&S&220&     &-18.11&0.41& 4467&100&3& 4330&1350& 18& 5.8&1.3\\[3pt]
163801263&HD 203922      & 8.50&60539.473&N& 75& 7602&-23.92&0.05& 3503& 30&2& 3830& 120& 78& 2.9&0.2\\[3pt]
170419024&HD 151860      & 9.01&55326.858&U&390& 6625&  3.45&0.07& 2265& 40&2& 2580& 170& 26& 4.2&0.3\\[3pt]
238659021&HD 8441        & 6.68&60686.361&N&160& 9205&  3.68&0.03&     &   & &   --&    & 64& 3.1&0.2\\[3pt]
251976407&HD 221568      & 7.55&60686.313&N&125& 9472& -8.33&0.12& 2433& 30&2& 2790& 250& 38&  --&\\[3pt]
301918605&HD 17330       & 7.11&60686.421&N&240&10250&-12.42&0.03&     &   & & 1130& 180& 72& 1.4&0.1\\[3pt]
341616734&HD 89069       & 8.42&60663.707&N&100& 9546& -9.23&0.10&     &   & & 3380& 120& 83& 4.3&0.2\\
         &               &     &60686.678&N&100&     & -9.17&0.05&     &   & & 3200&  90& 78& 4.8&0.1\\[3pt]
352787151&BD+35 5094     & 9.08&60686.350&N& 85& 6900& -6.81&0.05&     &   & &   --&    & 85& 2.0&0.2\\[3pt]
356088697&HD 76460       & 9.80&55227.707&F&270& 7110&      &    & 3607& 40&2& 3890& 260& 64& 4.0&0.3\\  
         &               &     &60038.395&S&210&     &  2.20&0.11& 3586& 50&2& 3350& 340& 60& 3.9&0.4\\[3pt]
369969602&HD 128472      & 9.87&60484.425&S&210& 7259&-16.94&0.15&     &   & & 1840& 570& 46& 3.3&0.4\\[3pt]
380607580&HD 119794      & 9.00&60704.562&S&145& 9347&-21.49&0.16& 5369& 40&2& 4800& 400& 47& 6.1&0.5\\[3pt]
403625657&HD 11187       & 7.94&60539.704&N& 65&10750&  5.73&0.06&     &   & &   --&    & 49&15.3&0.1\\[3pt]
405516045&HD 110274      & 9.47&54138.884&F&180& 7766&      &    & 3775& 40&3& 3580& 440& 39& 2.5&0.5\\
         &               &     &54141.828&F&100&     & -1.43&0.10& 3754& 40&3& 3350& 370& 42& 4.1&0.4\\
         &               &     &54172.906&U&150&     & -7.75&0.08& 5034& 30&2& 6320& 150& 33&  --&\\ 
         &               &     &54523.854&U&115&     & -2.97&0.08& 4698& 30&2& 6000& 170& 33&  --&\\
         &               &     &54535.815&U&170&     & -4.94&0.09& 4425& 30&2& 5670& 200& 33& 1.5&0.5\\  
         &               &     &60032.587&S&250&     & -1.49&0.14& 5243& 40&2& 5850& 260& 38& 3.2&0.4\\[3pt]
419916333&HD 117290      & 9.25&54139.763&F&160& 7271&      &    & 6439& 40&2& 7320& 330& 21&  --&\\
         &               &     &54141.866&F&130&     &-27.57&0.23& 6313& 40&2& 7540& 310& 22&  --&\\
         &               &     &54171.921&U&120&     &-26.97&0.16& 6334& 30&2& 7650& 260& 27&  --&\\
         &               &     &54515.870&U&250&     & -4.47&0.15& 4509& 30&2& 6550& 270& 27&  --&\\
         &               &     &60036.350&S&260&     &-36.87&0.20& 3020& 40&2& 2630& 630& 34& 2.2&0.7\\[3pt]  
430355895&BD+52 3124     &10.12&60540.656&N& 50& 9066&-10.94&0.07& 4237& 50&2& 4920& 170& 55& 2.3&0.4\\[3pt]
470837956&BD+61 2565     & 9.9 &60555.089&N&200& 8178&-17.89&0.06&     &   & & 1680& 140&111& 2.4&0.2\\[4pt]
\hline\\[-4pt]
\end{tabular*}
  \tablefoot{The stars are listed in order of increasing TIC (TESS
    Input Catalogue) number
  (Col.~1), with another ID (preferably the HD number) given in
  Col.~2 and the $V$ magnitude in Col.~3. The Heliocentric
  Julian Date (HJD) of mid-observation, 
  the instrument with which it was obtained (F\,=\,FEROS; 
  H\,=\,HARPS; N\,=\,HARPS-N; S\,=\,SALT-HRS; U\,=\,UVES), and the
  resulting S/N appear 
  in Cols.~4 to 6. The values of \Teff\ in Col.~7 are from the
  TIC (they are listed as published there even though effective
  temperatures cannot be determined with 1\,K precision). The
  following columns contain the  
  results of our measurements: the stellar radial velocity $v_{\rm
    r}$ and its uncertainty $\sigma(v_{\rm r})$ (Cols. 8 and 9; the
  entries were left blank when these values could not be determined);
  the 
  mean magnetic field modulus $\Bm$, its uncertainty $\sigma(\Bm)$,
  and the number $N_{\rm g}$ of Gaussians fitted to the \Feline\ line
  for its determination (Cols.~10 to 12); the mean quadratic magnetic
  field $\Bq$, its 
  formal uncertainty $\sigma(\Bq)$, and the number $N_{\rm l}$ of
  diagnostic lines from which it was derived (Cols.~13 to 15); the
  upper limit $(\vsi)_{\rm max}$ of the projected equatorial velocity
  and the formal uncertainty $\sigma(\vsi)$ of the \vsi\
  determination (Cols.~16 and 17). The values of the radial
  velocity and of the projected equatorial velocity are derived from
  analysis of the same sample of $N_{\rm l}$ lines as used for the $\Bq$
  determination. For spectra in which the \Feline\ line is not
  resolved into its magnetically split components, the entries in
  Cols.~8 to 10 are left blank; a dash (`--') in Col.~11 identifies those
  spectra in which the mean quadratic magnetic field is below the
  detection threshold, and in Col.~14, those spectra in which the
  rotational Doppler broadening is below the detection threshold. (For
TIC~93522454, the $\Bq$ value derived from the first UVES spectrum was
given as it is only marginally below the formal detection threshold
and as significant values of similar order were determined from the
other UVES observations of this star.)} 
\end{table*}

\afterpage{\clearpage}
\begin{figure*}
  \centering 
  \includegraphics[width=0.9\textwidth]{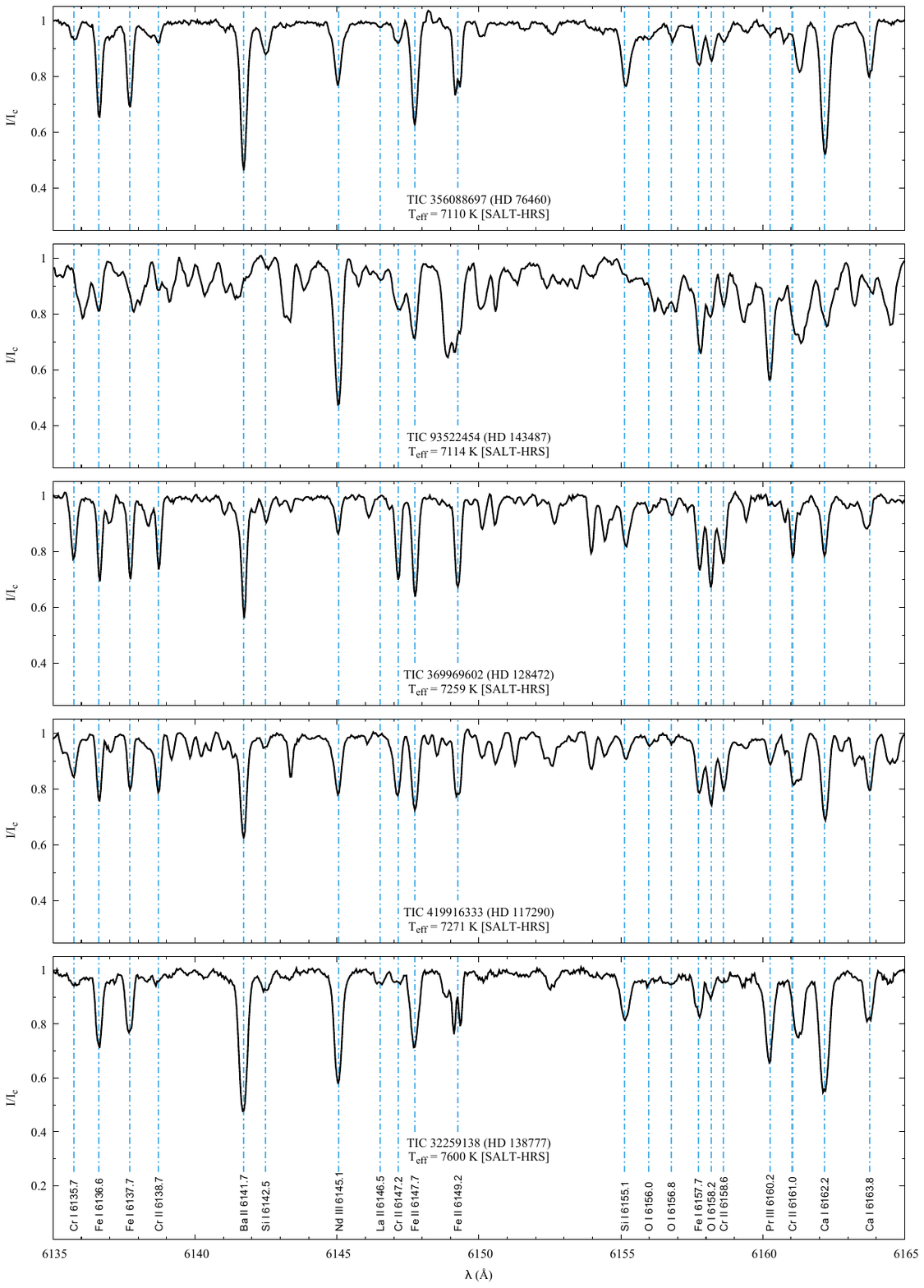}
  \caption{Portion of the spectrum of five sharp-lined ssrAp star 
    candidates. The wavelengths are in the laboratory reference frame.}
  \label{fig:spec6150_1}
\end{figure*}

\afterpage{\clearpage}
\begin{figure*}[p]
  \centering 
  \includegraphics[width=0.9\textwidth]{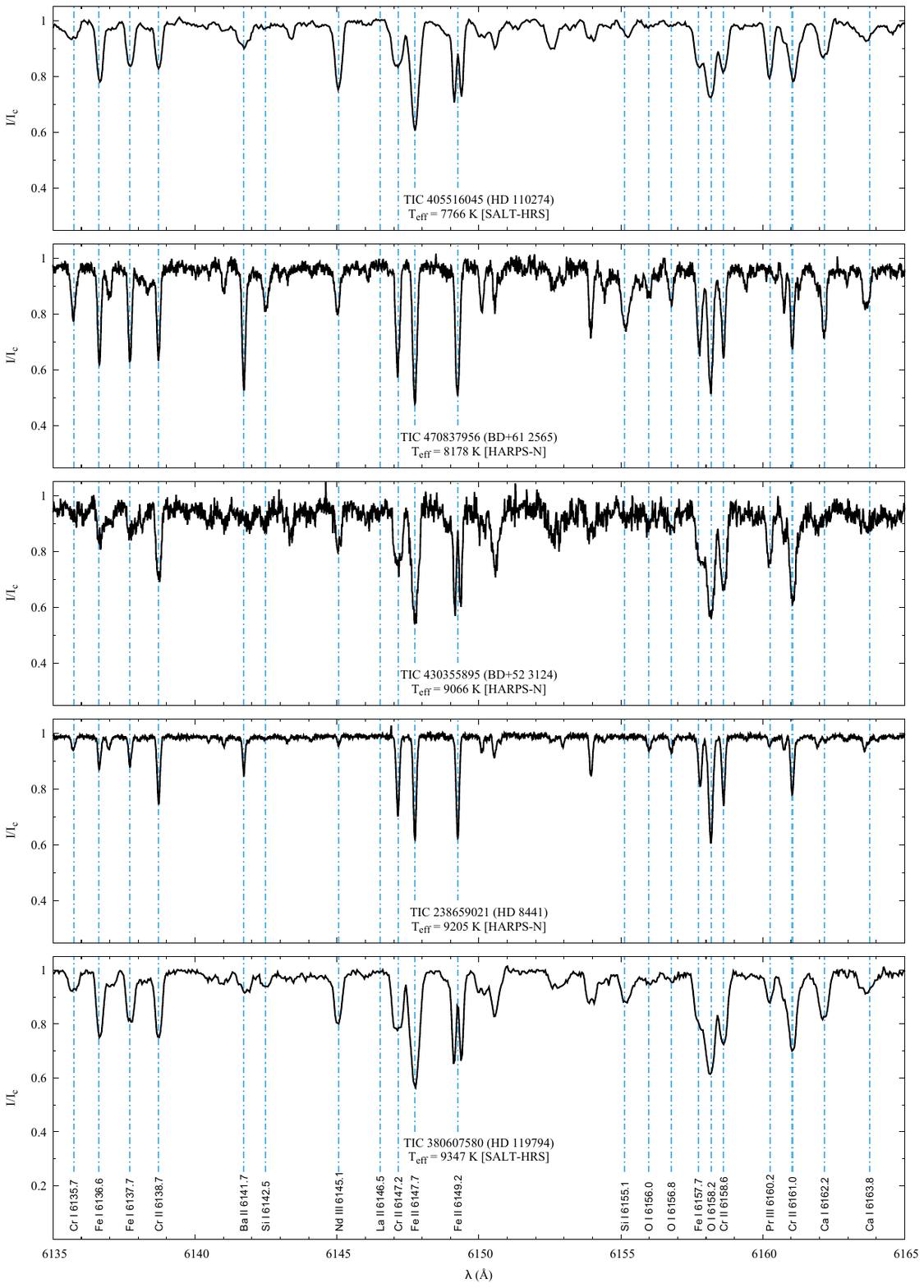}
  \caption{Portion of the spectrum of five sharp-lined ssrAp star 
    candidates. The wavelengths are in the laboratory reference frame.}
  \label{fig:spec6150_2}
\end{figure*}

\afterpage{\clearpage}
\begin{figure*}[t]
  \centering 
  \includegraphics[width=0.9\textwidth]{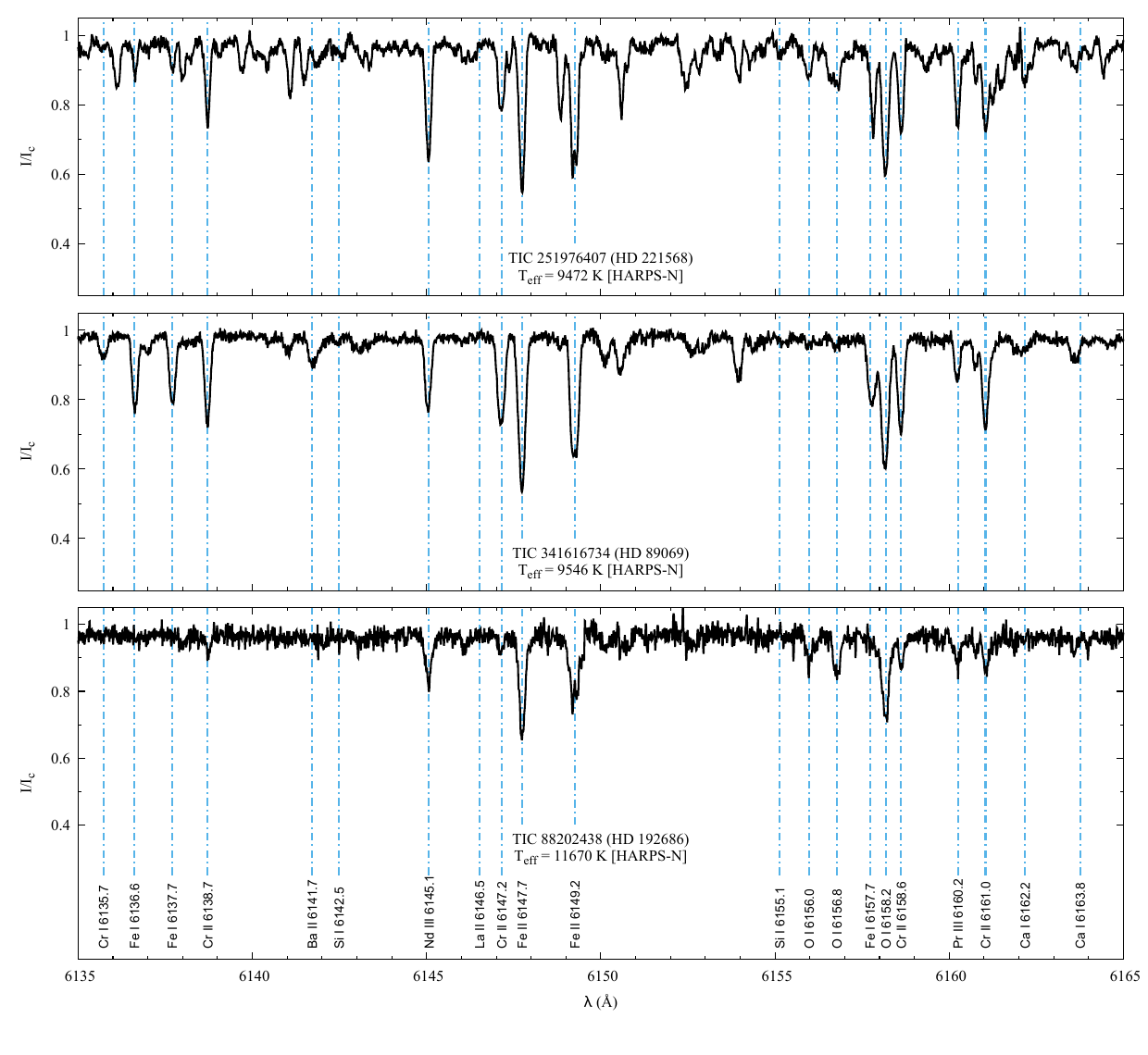}
  \caption{Portion of the spectrum of three sharp-lined ssrAp star 
    candidates. The wavelengths are in the laboratory reference frame.}
  \label{fig:spec6150_3}
\end{figure*}

\afterpage{\clearpage}
\begin{figure*}
  \includegraphics[width=\textwidth]{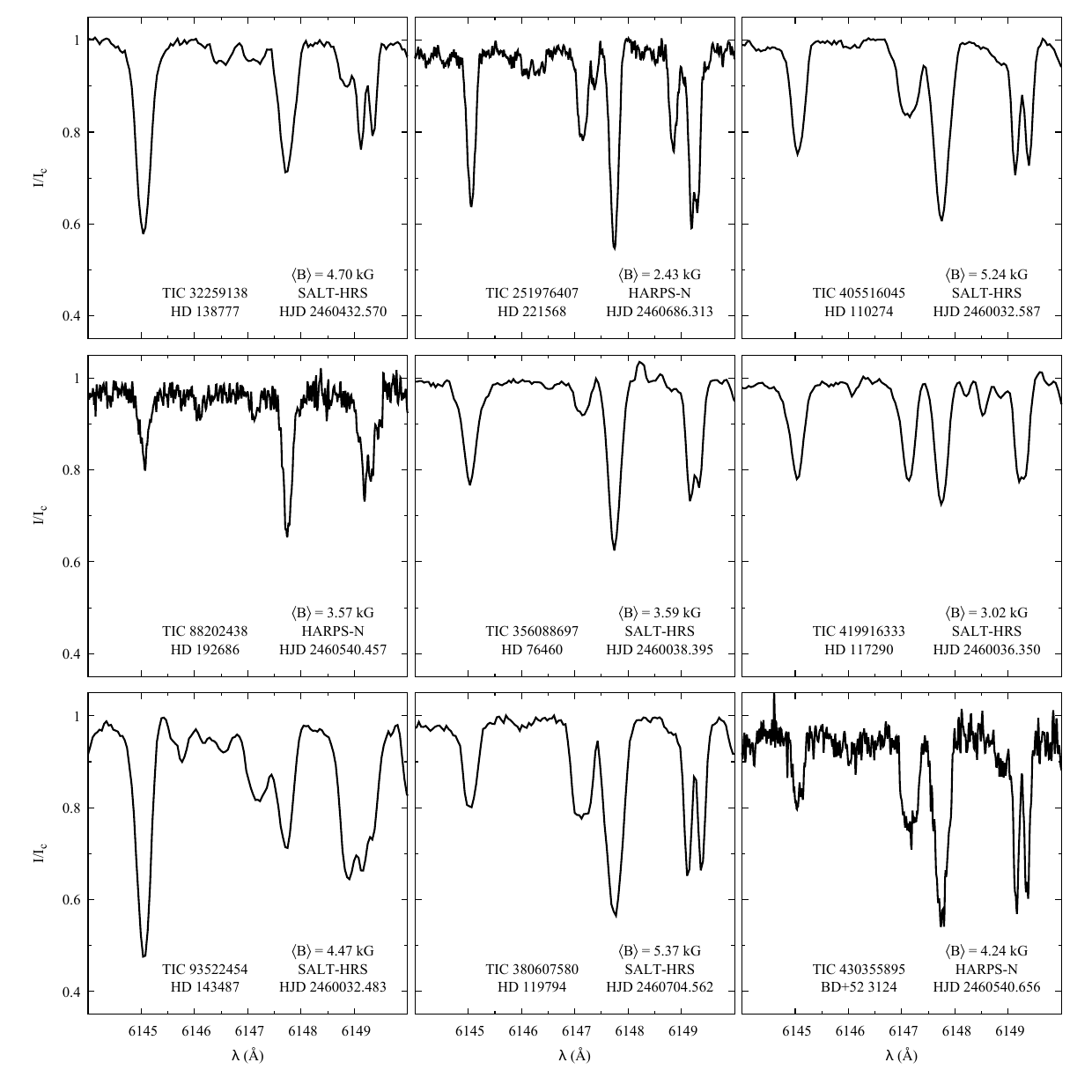}
  \caption{Portion of the spectrum of the nine Ap stars with resolved
    magnetically split lines that were not studied by
    \citet{2024A&A...691A.186M}. The wavelengths are in the
    laboratory reference frame. The main lines are Nd~{\sc
      iii}~$\uplambda\,6145.1$\,\AA, Cr~{\sc ii}~$\uplambda\,6147.2$\,\AA,
    Fe~{\sc ii}~$\uplambda\,6147.6$\,\AA, and \Feline; their Zeeman
    patterns were shown in Fig.~3 of
    \citet{2024A&A...691A.186M}. The doublet pattern of the
    \Feline\ line is of particular interest as it lends itself to 
    mostly approximation-free and model-independent determination of
    the mean magnetic field 
    modulus $\Bm$. However, in a fraction of the stars, its blue wing
    is affected to some extent by blends that may include 
    contributions from the Sm~{\sc ii}~$\uplambda\,6149.06$\,\AA\ line
    and/or from an unidentified line. This blend is exceptionally
    strong in HD~143487, which also features the strongest Nd~{\sc
      iii}~$\uplambda\,6145.2$\,\AA\ line of all the stars shown
    here. The benefit of the higher resolving power of HARPS-N over
    the lower one of SALT-HRS for separation of the magnetically split
    components of the \Feline\ line is also apparent in this figure.}
      \label{fig:spec6149}
  \end{figure*}

\afterpage{\clearpage}
\begin{figure*}[t]
  \centering
  \includegraphics[scale=0.81]{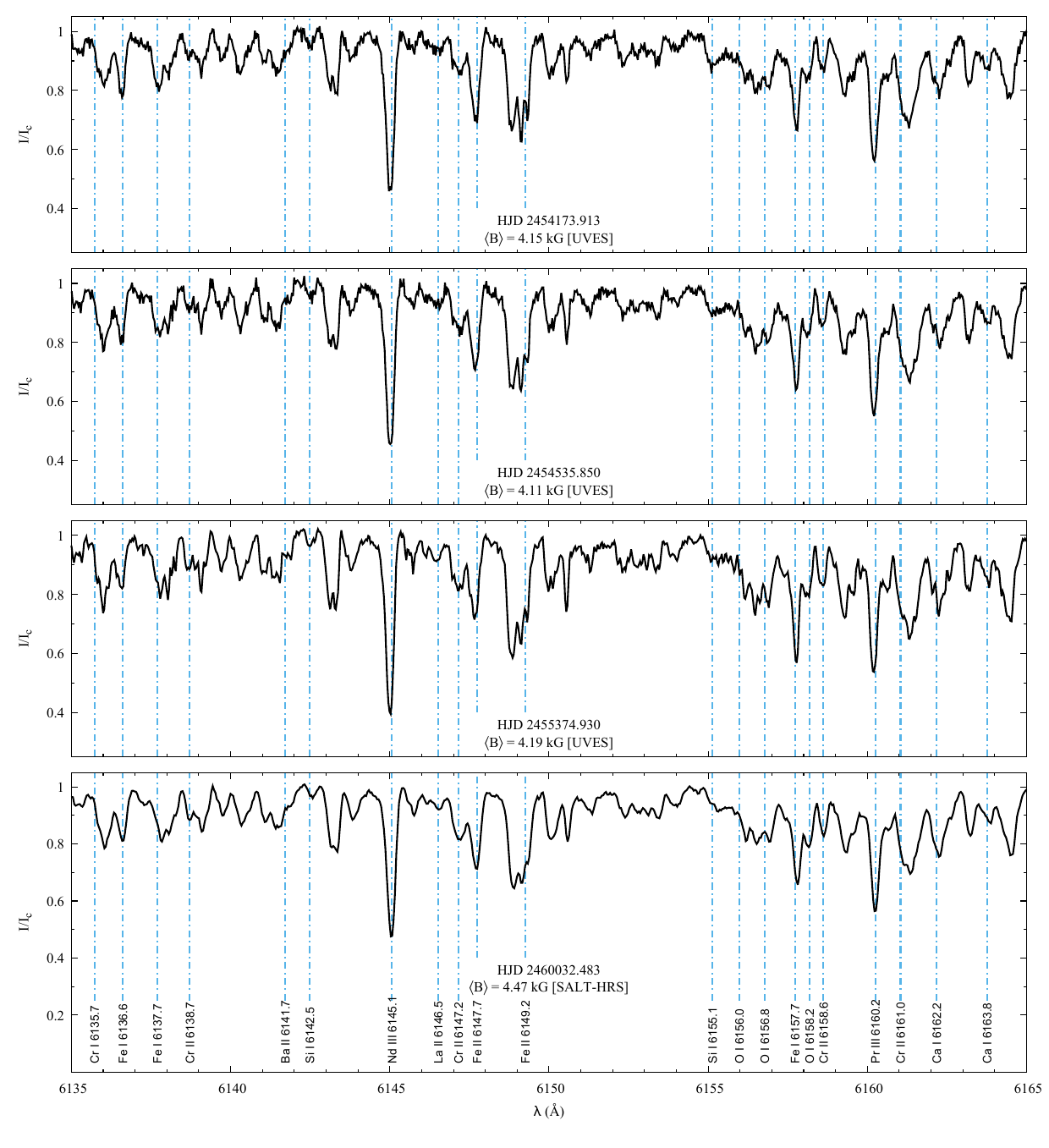}
  \caption{Portion of the spectrum of HD~143487 observed at 4
    epochs. The wavelengths are in the laboratory reference
    frame. The same lines as in Figs.~\ref{fig:spec6150_1} to
    \ref{fig:spec6150_3} are identified even though some
    may not be seen in the present spectra. The blend affecting the
    blue side of the \Feline\ line, which probably includes a
    contribution of the Sm~{\sc ii}~$\uplambda\,6149.06$\,\AA\ line in
    addition to an unidentified 
    line, is deeper than the Fe doublet in the lower (more recent) two
    spectra than in the upper two. The increase in depth of the
    Cr~{\sc ii}~$\uplambda\,6147.2$\,\AA\ line from the top to the bottom
    spectrum is also quite apparent by comparison with the
    neighbouring Fe~{\sc ii}~$\uplambda\,6147.7$\,\AA\ line. These
    relative intensity changes in line pairs are suggestive of
    variations occurring over time scales of years, which is
    consistent with super-slow rotation.}   
  \label{fig:hd143487_6150}
\end{figure*}

\afterpage{\clearpage}
\begin{figure*}[t]
  \centering
  \includegraphics[scale=0.81]{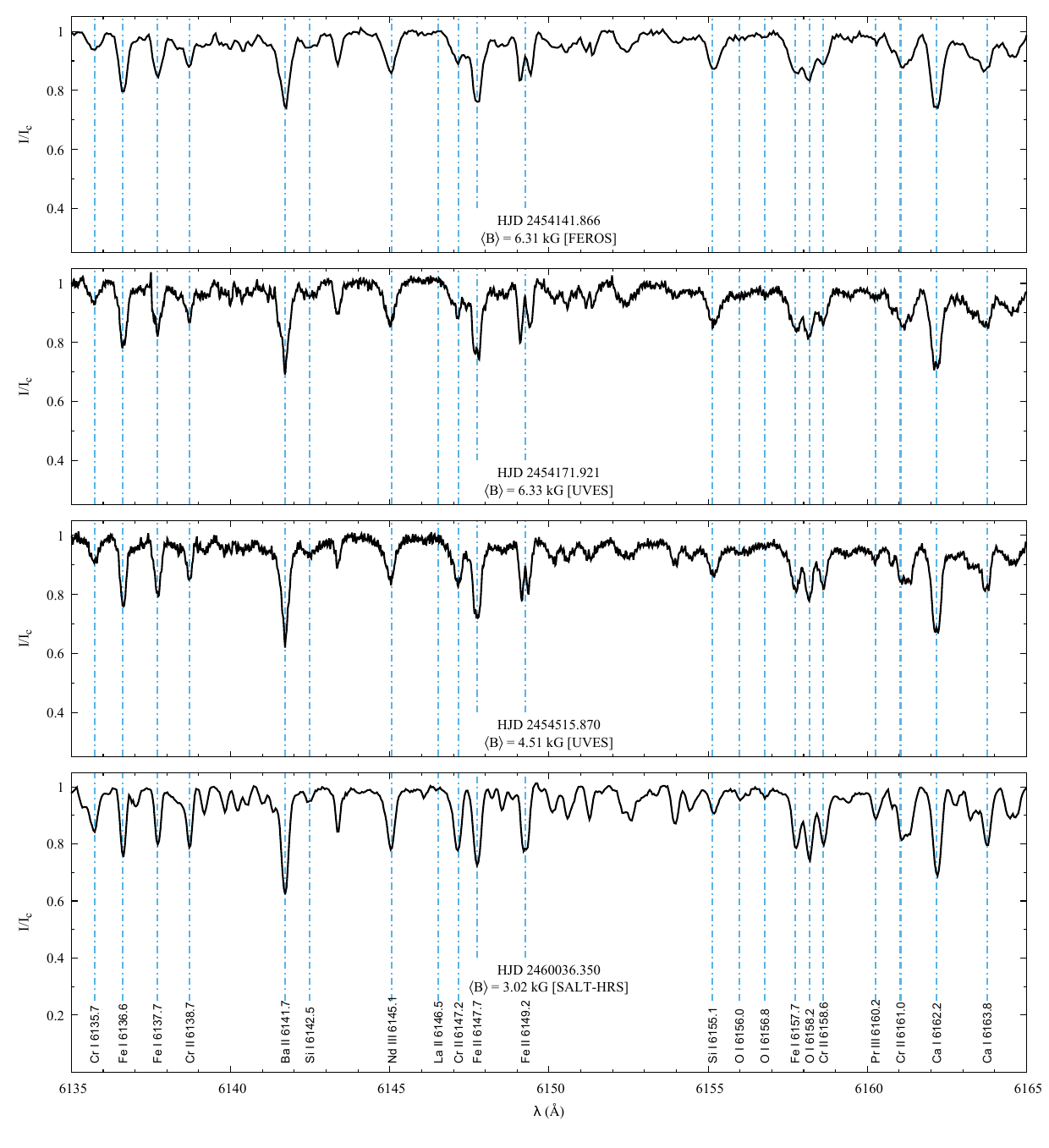}
  \caption{Portion of the spectrum of HD~117290 observed at 4
    epochs. The wavelengths are in the laboratory reference
    frame. The same lines as in Figs.~\ref{fig:spec6150_1} to
    \ref{fig:spec6150_3} are identified even though some
    may not be seen in the present spectra. The narrowing splitting of
    the \Feline\ line clearly shows 
    the decrease of the mean magnetic field modulus with time. The
  increase in intensity of the Cr lines from the first observation to
  the most recent one is particularly notable from comparison of the
  Cr~{\sc ii}~$\uplambda\,6147.2$\,\AA\ line with the neighbouring
  Fe~{\sc ii}~$\uplambda\,6147.7$\,\AA\ and $\uplambda\,6149.2$\,\AA\
  lines. Other Cr lines show similar variations. The strengthening of
  the Nd~{\sc iii}~$\uplambda\,6145.1$\,\AA\ 
  and Pr~{\sc iii}~$\uplambda\,6160.2$\,\AA\ with weakening $\Bm$ is
  also visible. Comparison of the top two spectra, obtained
  respectively with FEROS and UVES at epochs between which the star
  did not significantly vary allows one to assess the effect of the
  resolution on the line profile appearance.} 
  \label{fig:hd117290_6150}
\end{figure*}

\end{appendix}
\end{document}